\documentclass[fleqn,usenatbib]{mnras}

\usepackage{newtxtext,newtxmath}
\usepackage[T1]{fontenc}
\usepackage{subfigure}

\DeclareRobustCommand{\VAN}[3]{#2}
\let\VANthebibliography\thebibliography
\def\thebibliography{\DeclareRobustCommand{\VAN}[3]{##3}\VANthebibliography}

\usepackage{graphicx}	
\usepackage{amsmath}	
\usepackage{xcolor}
\usepackage{soul}

\newcommand{\FML}[1]{\textsc{FML}-COLA{#1}}
\newcommand{\taka}{Takahashi {\it et al.}}

\title[Weak Lensing Map Generation in MG with COLA]{Fast Generation of Weak Lensing Maps in Modified Gravity with COLA}

\author[]{
Sophie Hoyland\thanks{E-mail: \url{sophie.hoyland@port.ac.uk}},$^{1}$
Hans A. Winther,$^{2}$
Daniela Saadeh,$^{1}$
Kazuya Koyama,$^{1}$
and Albert Izard
\\
$^{1}$Institute of Cosmology and Gravitation, University of Portsmouth, Dennis Sciama Building, Burnaby Road, Portsmouth, PO1 3FX, United Kingdom \\
$^{2}$Institute of Theoretical Astrophysics, University of Oslo, Blindern 0315, Oslo, Norway \\
}

\date{Accepted XXX. Received YYY; in original form ZZZ}
\pubyear{\the\year{}}

\begin{document}
\label{firstpage}
\pagerange{\pageref{firstpage}--\pageref{lastpage}}
\maketitle

\begin{abstract}

Accurate predictions of weak lensing observables are essential for understanding the large-scale structure of the Universe and probing the nature of gravity. In this work, we present a lightcone implementation to generate maps of the weak lensing convergence field using the COmoving Lagrangian Acceleration (COLA) method. The lightcone is constructed in spherical shells from the source to the observer following an onion representation of the Universe. We validate the COLA-generated convergence maps in General Relativity by comparing five statistics to those of maps obtained with publically available high-resolution {\it N}-body simulations: the power spectrum, bispectrum, probability distribution function, peak counts and Minkowski functionals. The convergence power spectrum is accurate to within $5\%$ up to $\ell\sim500$ and to within $10\%$ up to $\ell\sim750$, confirming the accuracy of this method on both linear and non-linear scales. For the probability distribution function, peak counts and Minkowski functionals, we determine the map pixel resolution required for COLA to capture the statistical features of the {\it N}-body convergence maps. Our validation tests provide a baseline for the convergence map specifications at which we can trust COLA for each statistic considered. Using these map specifications, we extend our analyses to two representative theories of Modified Gravity, and demonstrate their imprints on the five convergence statistics considered. This work represents a step towards precise weak lensing predictions under both General Relativity and Modified Gravity with reduced computational cost, providing a robust framework to explore the nature of gravity using field-level inference.

\end{abstract}


\section{Introduction}\label{sec:intro}


The large-scale distribution of matter in the Universe provides a unique opportunity to test one of the most fundamental assumptions of modern cosmology: that gravity is described by Einstein's theory of General Relativity (GR) across all cosmological scales. The standard $\Lambda$CDM model is a remarkable achievement of modern cosmology, providing an excellent fit to observations across multiple scales, from Solar System dynamics \citep{will_confrontation_2014} to the propagation of gravitational waves \citep{abbott_gw170817_2017, creminelli_dark_2017, collaboration_gravitational_2017}. However, tensions have emerged such as discrepancies between CMB and lensing measurements of the growth of structure, $\sigma_8$ (see \citealt{abdalla_cosmology_2022} for a review), and CMB and supernovae measurements of the expansion rate of the Universe, $H_0$ (see \citealt{valentino_realm_2021} for a review). These tensions have motivated models beyond $\Lambda$CDM and are an incentive to test gravity on cosmological scales (see \citealt{koyama_cosmological_2016} for a review). Modified Gravity (MG) theories \citep{clifton_modified_2012} can introduce additional degrees of freedom that change the growth of structures and the underlying geometry of spacetime, therefore providing a framework for cosmological tests of gravity, particularly in the non-linear regime, where gravitational collapse leads to the formation of galaxies, galaxy clusters and other large-scale structures. 

Although GR has been rigorously tested within the Solar System \citep{will_confrontation_2014, baker_novel_2021}, its validity on the largest scales of the Universe remains poorly tested \citep{koyama_cosmological_2016}. Two of the most promising probes for constraining deviations from GR include galaxy clustering and weak gravitational lensing \citep{albrecht_report_2006, weinberg_observational_2013, munshi_galaxy_2016}. Weak lensing is the distortion of light from distant sources due to the gravitational field of the intervening matter, regardless of whether it is luminous or dark (see \citealt{bartelmann_weak_2001} for a review). The observed lensing effect is therefore sensitive to the growth of matter fluctuations and the Universe's expansion history. Stage IV surveys such as Euclid \citep{collaboration_euclid_2024} and the Vera Rubin Observatory \citep{ivezic_lsst_2019} promise to deliver high-precision weak lensing maps of increasing volume, providing a large-scale cosmological experiment that will allow us to distinguish between different theories of dark energy and MG.

Crucially, weak lensing probes the non-linear regime of structure formation, where gravitational collapse drives the growth of matter perturbations beyond the linear approximation \citep{frenk_nonlinear_1983, bernardeau_large-scale_2002}. This introduces mode coupling, which arises in a highly non-Gaussian matter distribution. Under the Born approximation, the weak lensing convergence field is the projection of this matter distribution along the line of sight. To fully extract the information contained within non-Gaussian fields, we must consider statistics beyond traditional summary statistics, such as the power spectrum (the two-point correlation function, 2PCF), which only capture the Gaussian properties of the field. Higher-order statistics, such as the bispectrum ({\it e.g.,} \citealt{dodelson_weak_2005, 
cooray_weak_2000, munshi_weak_2020}), peak counts ({\it e.g.,} \citealt{marian_cosmology_2009, harnois-deraps_testing_2015, martinet_constraining_2015, kacprzak_cosmology_2016, martinet_kids-450_2018}), probability distribution function (PDF; {\it e.g.,} \citealt{barthelemy_nulling_2020, thiele_accurate_2020,  boyle_nuw_2021, einasto_evolution_2021}) and Minkowski functionals (MFs; {\it e.g.,} \citealt{schmalzing_minkowski_1995, kratochvil_probing_2009, petri_emulating_2015, parroni_going_2020}), offer complementary insights into the growth of structure in the non-linear regime under different models of gravity \citep{ling_distinguishing_2015, shirasaki_imprint_2017, giocoli_weak_2018-1, cataneo_matter_2022, gough_one-point_2022, davies_constraining_2024, jiang_minkowski_2024}. Ultimately, any summary statistic is a lossy compression of the full-field information: several works have shown that field-level inference based on maps achieves superior cosmological constraints compared to
two-point analyses \citep{leclercq_accuracy_2021, boruah_map-based_2023, porqueres_field-level_2023,   zhou_accurate_2023, mancini_field-level_2024}.

To fully leverage the statistical power of weak lensing data, it is essential to accurately model non-linear structure growth under different cosmological models. Predicting large-scale structure properties on the non-linear scales allows us to compare mocks to observational datasets, and is critical for placing constraints on particular theories through covariance and error estimation. However, modelling the non-linear growth of structure demands computationally expensive {\it N}-body simulations, while accurate error estimation requires a large number of mock catalogues \citep{hartlap_why_2007}. This has motivated the development of faster, approximate methods such as the COmoving Lagrangian Acceleration (COLA) approach \citep{tassev_solving_2013}, which allows for faster predictions of dark matter structures at late times while maintaining the accuracy of full {\it N}-body simulations on large-scales at the expense of less accuracy on small scales \citep{tassev_scola_2015, winther_cola_2017, fiorini_fast_2021, ding_fast_2024}. The COLA method has been extended to beyond-$\Lambda$CDM cosmologies \citep{wright_cola_2017, wright_texttthi-cola_2023, fiorini_studying_2022, gordon_modeling_2024}, allowing for a self-consistent comparison of large-scale structure observables predicted under GR and MG. 

In this work, we present a new implementation of the COLA method to predict the weak lensing convergence field by constructing a lightcone between the source redshift and the observer at $z=0$. In particular, this lightcone method is implemented into the \FML{} (Fourier-Multigrid Library) library\footnote{\url{https://github.com/HAWinther/FML/}}, which easily allows us to use the COLA method for many MG scenarios. 

The remainder of this paper is organised as follows: in Section~\ref{sec:WL}, we present a detailed description of the implementation in the \FML{} library, specifically an overview of the COLA method, the lightcone generation, and the convergence map construction. In Section~\ref{sec:MG}, we provide an overview of the MG theories considered within this work, namely $f(R)$ and normal-branch Dvali-Gabadadze-Poratti (nDGP) gravity. The five statistics we use to analyse the predicted convergence field -- the power spectrum, bispectrum, PDF, peak counts and MFs -- are defined in Section~\ref{sec:stats}. In Section~\ref{sec:validation}, we demonstrate the accuracy of the convergence field predicted by \FML{} in GR through comparison to high-resolution maps generated by full {\it N}-body simulations. Additionally, we determine the optimal convergence map specifications required for each summary statistic. We then extend our analysis to two theories of MG in Section~\ref{sec:extension_to_MG}, and explore the statistical signatures of MG in the weak lensing convergence field. Finally, a discussion of this work is provided in Section~\ref{sec:discussion}.

\section{Weak Lensing Maps}\label{sec:WL}

The most precise way to generate weak lensing maps is to perform ray-tracing simulations \citep{hilbert_ray-tracing_2009}, which directly compute the distortion and magnification effects by propagating light rays from the source to the observer. Whilst powerful, this approach is computationally expensive, and such simulations often focus on small patches. A faster approach is to create a lightcone \citep{evrard_galaxy_2002, fosalba_onion_2008, izard_ice-cola_2018, arnold_modified_2019}, where particles are output from an {\it N}-body simulation as soon as they exit the lightcone of an observer located at $z=0$ (see Section~\ref{sec:WL_lightcone} below). The lightcone output is then used to create mass maps ("onion shells"), which are subsequently summed up with the appropriate lensing weight. This is the approach we take. This procedure is based on the Born approximation that the perturbations to the light path induced by gravitational lensing are negligible. 

Ray tracing is necessary to achieve percent level accuracy in higher-order statistics \citep{ferlito_ray-tracing_2023}. However, a lightcone approach is sufficient, and better suited, for our purpose of fast generation of weak lensing maps using the approximate COLA method. This section details the methods employed in this work to create weak lensing maps, including the COLA method, the lightcone construction and weak lensing map generation.

\subsection{COLA method}\label{sec:WL_COLA_method}

Within this work, we have run Particle-Mesh (PM) lightcone {\it N}-body simulations employing the COmoving Lagrangian Acceleration (COLA) method \citep{tassev_solving_2013}. COLA provides a fast, approximate simulation approach for modelling non-linear structure formation, by combining aspects of full {\it N}-body simulations and Lagrangian Perturbation Theory (LPT) to evolve particles.

In COLA, the particle positions are decomposed as:
\begin{align}
    {\bf x}(t) = {\bf x}_{\rm LPT}(t) + {\bf x}_{\rm res}(t),
\end{align}
where ${\bf x}_{\rm LPT}$ is the particle's trajectory predicted by LPT and ${\bf x}_{\rm res}$ is a residual which is not necessarily assumed to be small. The equations evolved by an {\it N}-body code:
\begin{align}
    \frac{d{\bf x}}{dt} &= {\bf v},\\
    \frac{d{\bf v}}{dt} &= -\nabla\Phi,
\end{align}
can then be written as
\begin{align}
    \frac{d{\bf x}_{\rm res}}{dt} &= {\bf v}_{\rm res},\\
    \frac{d{\bf v}_{\rm res}}{dt} &= -\nabla\Phi - \frac{d^2{\bf x}_{\rm LPT}}{dt^2}.
\end{align}

The equations for $({\bf x}_{\rm res},{\bf v}_{\rm res})$ are now in the same form as those for $({\bf x},{\bf v})$, with the exception that there is an additional "force", $\frac{d^2{\bf x}_{\rm LPT}}{dt^2}$, due to the choice of frame: in a COLA simulation, we are effectively evolving the particles in a frame following their LPT trajectories. At early times and for large scales, ${\bf v}_{\rm res}$, and thereby ${\bf x}_{\rm res}$, is very small, and this allows us to take very large timesteps whilst retaining accuracy on the largest scales\footnote{For simulations without COLA, using large timesteps typically leads to power-loss on all scales.}. Other methods have been developed that have the same advantage, namely time-stepping algorithms that are designed to preserve 1LPT or 2LPT trajectories, regardless of the timestep (see \citealt{mcewen_fast-pt_2016, list_perturbation-theory_2024}).

The additional information we need, ${\bf x}_{\rm LPT}$, follows from LPT:
\begin{align}
{\bf x}_{\rm LPT}({\bf q},t) = {\bf q} + {\bf\Psi}({\bf q},t),    
\end{align}
where ${\bf\Psi}$ is the so-called displacement field as function of time and the initial (Lagrangian) position ${\bf q}$ of the particle. To first order in LPT, the displacement field is simply given by the Zel'dovich approximation \citep{shandarin_large-scale_1989}, ${\bf\Psi}= {\bf\Psi}_{\rm ini}({\bf q}) D(t)$, where $\nabla_q{\bf \Psi}({\bf q},t_{\rm ini}) = -\delta_{\rm ini}$, $D(t)$ is the linear growth-factor and $\delta_{\rm ini}$ is the initial overdensity field. ${\bf \Psi}$ is computed when generating the initial conditions of the simulation, so this additional data required to use COLA essentially comes for free (though extra storage is needed to hold this information). 

Any time-stepping algorithm can be used in conjunction with the COLA method: in this work, we use the leapfrog algorithm, whereby particle positions and velocities are updated in three interleaved steps called `kick-drift-kick'. In 'kick' steps, velocities are updated using accelerations, whereas in 'drift' steps, positions change following the velocities. We will make reference to these steps in Section~\ref{sec:WL_lightcone} when detailing the lightcone construction.

COLA can accurately capture the growth of structures down to fairly non-linear scales ($\sim 1\%$ agreement in $P(k)$ up to $k \sim 1 h$ Mpc$^{-1}$ \citep{izard_ice-cola_2016, winther_cola_2017}), yet the computational cost is typically reduced by a factor $\mathcal{O}(100-1000)$ compared to traditional {\it N}-body simulations. The PM technique is implemented on a fixed grid, where the size of the grid defines the force resolution of the simulation. The COLA accuracy therefore depends on the number of grids and the number of timesteps, meaning there is a trade-off between speed and accuracy. Beyond scales of $k\sim1 h$ Mpc$^{-1}$, baryonic effects become important and COLA, like any dark matter-only simulation, has limited accuracy. However, when studying alternative models, taking the ratio of two power spectra ({\it i.e.} $P / P_{\Lambda \rm CDM}$) cancels out some of this inaccuracy, allowing us to extend the result to even smaller scales \citep{brando_enabling_2022}. 

\subsection{Lightcone construction}\label{sec:WL_lightcone}

To simulate weak lensing observables, lightcones are constructed on-the-fly during the COLA simulations, following the procedure presented in \cite{izard_ice-cola_2018} and \cite{fosalba_onion_2008, fosalba_mice_2015}. We assume the Born approximation, whereby photons remain unperturbed as they propagate through each lens plane. This allows us to discretise the lightcone into spherical concentric shells separated by constant intervals of the scale factor, $\Delta a$, following the onion-like representation of the universe \citep{fosalba_onion_2008}. This approximation allows for a considerable speed-up compared to ray-tracing techniques, which need to constantly compute the geometry and deflection angles as they follow photons along their perturbed path from the source to the observer. 

At each timestep of the COLA simulation, the positions and velocities of the dark matter particles are updated according to their LPT trajectories, as described in Section~\ref{sec:WL_COLA_method}. After each set of drift and kick operators, the lightcone routine is executed to construct the shell at a given radius, and to compute the time at which a photon crosses the lightcone. This is called the crossing time, $t_c$, and is computed according to:
\begin{equation}
    t_c = \frac{c}{c + v_{\rm rad}}(t_p - t_D) + t_D,
\end{equation}
where $v_{\mathrm{rad}}$ is the radial velocity of the particle, $t_p$ is the time when the particle crosses the shell boundary and $t_D$ is the time of the previous drift operator. To ensure accurate tracking of particles near the shell boundaries, a spherical buffer is implemented around the lightcone. This buffer assumes that the maximum particle speed does not exceed $2\%$ of the speed of light ({\it i.e.} $v/c=0.02$). Particles that do not belong to the shell volume or the buffer are omitted from the lightcone data.

We can expand the simulated volume by means of box replicas, where particle are copied and their coordinates shifted by a box side length along Cartesian axes, placed around the observer. Periodic boundary conditions ensure a continuous dark matter distribution at the box boundaries. With no box replicas, and with the observer placed at (0,0,0), the observer only sees 1/8th of the sky. However, box replicas can be implemented to cover the full-sky. The replicated patterns are justified because we observe different projection angles at different redshifts and in different box replicas, meaning that each octant on the sky is nearly statistically independent. The comoving size of the lightcone is then determined by the initial redshift at which it switches on, corresponding to the source redshift, $z_{*}$, and the number of box replicas used. We note that the observer's position can be changed to create different realisations of the lightcone.

\subsection{Convergence map generation}\label{sec:WL_map_construction}

At each timestep, the dark matter density field in each onion shell of the lightcone is projected onto 2D sky maps using the \textsc{HEALPix} scheme \citep{gorski_healpix_2005}. Map resolution is determined by the parameter $N_{\rm side}$, with the total number of pixels in the map given by $N_{\rm pix}=12N_{\rm side}^2$. We allow for either a nested or ring pixelisation configuration, with nested being the default.

The weak lensing convergence field, $\kappa(\boldsymbol{\theta})$, is defined as the weighted integral of the matter density contrast along the line of sight:
\begin{equation}\label{eq:kappa}
    \kappa(\boldsymbol{\theta}, \chi_*) = \int_0^{\chi_*} W(\chi) \delta_m(\boldsymbol{\theta}, \chi) d\chi,
\end{equation}
where $\chi$ is the radial comoving distance, $\chi_*$ is the source distance and $\delta_m(\boldsymbol{\theta}, \chi)$ is the matter density contrast as a function of angular position, $\boldsymbol{\theta}$, and $\chi$. $W(\chi)$ is a lensing weight function that encodes the geometric properties of the source and the observer, defined as:
\begin{equation}
    W(\chi) = \frac{3H_0^2\Omega_m}{2c^2} \int_0^{\chi_*} \frac{(\chi_* - \chi)\chi}{a \chi_*} d\chi,
\end{equation}
where $H_0$ is the Hubble constant, $\Omega_m$ is the present-day matter density parameter, $c$ is the speed of light and $a$ is the scale factor.

To compute the convergence map in each radial bin, the integral in Eq.~\ref{eq:kappa} is discretised as follows \citep{fosalba_onion_2008}: 
\begin{equation}\label{eq:kappa_binned}
\kappa( \boldsymbol{\theta}_i, \chi_* ) = \frac{3 H_0^2 \Omega_m}{2 c^2} \sum_{j}^{\chi_j < \chi_* } \delta( \boldsymbol{\theta}_i, \chi_j ) \frac{(\chi_* - \chi_j) \chi_j}{a_j\ \chi_*} \Delta \chi_j,
\end{equation}
where $\Delta \chi_j$ is the width of the bin. The convergence field is computed on the fly at each timestep of the COLA run, integrating contributions from high redshift to low redshift. 

Each realisation of the lightcone COLA run produces particle data in Gadget format \cite{springel_cosmological_2005} and \textsc{HEALPix} maps at multiple timesteps. The angular power spectrum of $\kappa$ is also computed on-the-fly via the spherical harmonic decomposition of the \textsc{HEALPix} maps obtained through Eq.~\ref{eq:kappa_binned}.

\section{Modified Gravity}\label{sec:MG}

In this section, we provide a brief overview of the two theories of MG we consider within this work: $f(R)$ and nDGP. These are two representative examples of theories that include screening mechanisms, which suppress the enhancement of gravity on smaller scales where deviations from GR have been tightly constrained, such as within the Solar System. Theories that include such screening mechanisms can allow for a large-scale enhancement of gravity, while providing consistency with local observations, making them viable theories to test with current and future cosmological surveys. 

\subsection{$f(R)$}
Within $f(R)$ gravity, the Ricci scalar, $R$, in the standard Einstein-Hilbert action is replaced by a non-linear function of $R$, $f(R)$ \citep{hu_models_2007}. The modified action is given by
\begin{equation}
    S = \int d^4x \sqrt{-g}\left[\frac{1}{16\pi G} (R + f(R)) + \mathcal{L}_m \right],
\end{equation}
where $g$ is the determinant of the metric tensor, $G$ is the gravitational constant, and $\mathcal{L}_m$ is the Lagrangian of the matter fields. The function $f(R)$ introduces an additional scalar degree of freedom, which acts as a dynamical field, therefore introducing a fifth force between matter. We consider a simple model given by \citep{hu_models_2007}:
\begin{equation}
    f(R) = -2 \Lambda + f_{R0}\frac{R_0^2}{R},
\end{equation}
where $\Lambda$ is a cosmological constant and the present day value of this scalar field is denoted $-f_{R0}$. The GR limit is that where $f_{R0} \rightarrow 0$. $f(R)$ gravity is screened via the Chameleon mechanism \citep{khoury_chameleon_2004}, which depends on the gravitational potential of the local matter field and leads to a decoupling of the scalar field in high-density regions. We see a scale-dependent enhancement of the growth of linear density perturbations, while on non-linear scales the Chameleon mechanism ensures that GR is recovered locally to provide consistency with observational tests.

\subsection{nDGP}
The normal-branch Dvali-Gabadadze-Porrati (nDGP) model \citep{dvali_4d_2000} is motivated by higher-dimensional theories, where matter and radiation are confined to a 4D brane in a 5D Minkowski spacetime, while gravity is free to propagate from the 5D bulk. 
In the normal branch DGP model, as opposed to the self-accelerating branch, dark energy or a cosmological constant is still required to explain the accelerated expansion of the Universe. 

The enhancement of gravity in nDGP is suppressed by the Vainshtein screening mechanism, which switches on within a characteristic radius determined by the object's mass and cross-over scale, $r_c$. This screening mechanism ensures that any deviations from GR are reconciled in high-density regions.

\section{Convergence field statistics}\label{sec:stats}
The weak lensing convergence field, $\kappa$, contains valuable information about the underlying distribution of matter, particularly at the nonlinear scales. $\kappa$ is a highly non-Gaussian field as a result of mode coupling that occurs under non-linear structure formation, and therefore different statistics will be sensitive to difference scales and features.

Higher-order statistics are determined by non-Gaussian features, unlike the standard two-point correlation function (2PCF) or power spectrum.
A combination of the 2PCF and higher-order statistics will ensure that more information is extracted from the convergence field data. For example, the Weak Lensing Higher-order Statistics comparison project \citep{euclid_collaboration_euclid_2023-2} examined the constraining power of ten different higher-order statistics on Euclid-like mocks, and found that each individual higher-order statistic outperforms the 2PCF by a factor of two, while combining multiple higher-order statistics yields an improvement of up to 4.5 times that of the 2PCF alone. This emphasizes the role of higher-order statistics in capturing the intrinsically non-Gaussian nature of weak lensing signals and minimizing information loss.

In this section, we define the five statistics considered within our analyses: the power spectrum and bispectrum, the PDF, peak counts and MFs. 

\subsection{Power spectrum}\label{sec:stats_power_spectrum}

The power spectrum of the convergence field, $C_{\ell}^{\kappa\kappa}$, describes the amplitudes of fluctuations as a function of angular separation. 
Under the flat-sky approximation, it is defined as:
\begin{equation}
\langle{\kappa}(\boldsymbol{\ell}){\kappa}(\boldsymbol{\ell}')\rangle =
    (2\pi)^2 \delta_D(\boldsymbol{\ell} - \boldsymbol{\ell}')C^{\kappa\kappa}(\ell),
\end{equation}
where $\delta_D$ is the Dirac delta function, $\boldsymbol{\ell}$ is a 2D wave vector of modulus $\ell$, and angular brackets indicate the ensemble average over statistical realisations of the convergence signal. The angular multipole $\ell$ is inversely related to the angular scale in real space, $\theta$, through $\theta \sim \pi / \ell$. 

Given that the weak lensing convergence field is a projection of the three-dimensional density fluctuations along the line of sight, $C_{\ell}^{\kappa\kappa}$ can be expressed as an integration of the matter power spectrum, $P_{\delta}(k,z)$, weighted by a radial lensing kernel. The Limber approximation simplifies this projection by assuming that matter fluctuations, $\delta$, occur on scales that are much smaller than the lensing weight function. $C_{\ell}^{\kappa\kappa}$ can then be expressed as:
\begin{equation}\label{eq:C(ell)}
    C_{\ell}^{\kappa\kappa}(\chi_*) = \int_0^{\chi_*} \frac{W(\chi)^2}{r(\chi)^2} P_{\delta}( k, z(\chi)) d\chi, 
\end{equation}
where $\chi$ is the comoving distance to redshift $z$, $\chi_*$ is the comoving distance to the source redshift, $z_*$, and $r(\chi)$ is the comoving angular diameter distance. The 3D matter power spectrum is evaluated at the wavenumber $k = \ell / \chi$, and $W(\chi)$ is the lensing weight function, defined as:

\begin{equation}
    W(\chi_*) = \frac{3 H_0^2\Omega_{m}}{2 c^2} \frac{r(\chi_* - \chi) r(\chi) }{r(\chi_*)}(1 + z(\chi) ).
\end{equation}
The weight function $W(\chi)$ peaks at roughly half the comoving distance to the source object, and therefore this is where most of the lensing signal occurs. 

Predictions of the convergence power spectrum in MG theories such as $f(R)$ show that the most significant departures from GR are in the non-linear regime (see \citealt{ling_distinguishing_2015}). The convergence power spectrum can probe a range of scales from linear to non-linear regimes. 

\subsection{Bispectrum}\label{sec:stats_bispectrum}
Under the flat sky approximation, the bispectrum is defined as follows: 
\begin{equation}
\langle {\kappa}(\boldsymbol{\ell_1}){\kappa}(\boldsymbol{\ell_2}){\kappa}(\boldsymbol{\ell_3})\rangle= 
    (2\pi)^3\delta_D(\boldsymbol{\ell}_1 + \boldsymbol{\ell}_2 + \boldsymbol{\ell}_3)B(\ell_1, \ell_2, \ell_3),
\end{equation}
where $\ell_1, \ell_2$ and $\ell_3$ form a closed triangle $\boldsymbol{\ell}_1+\boldsymbol{\ell}_2+\boldsymbol{\ell}_3 =0$. The bispectrum depends on the configuration of these triangles, which can probe different physical scales and processes. For example, equilateral configurations ($\ell_1 = \ell_2 = \ell_3$) are sensitive to the non-linear regime, squeezed configurations ($\ell_1 \ll \ell_2 \approx \ell_3$) probe interactions between large and small scales, and folded configurations ($\ell_1 + \ell_2 \approx \ell_3$) can be sensitive to non-linear dynamics \citep{scoccimarro_bispectrum_1999, bernardeau_large-scale_2002, baldauf_primordial_2011, sefusatti_accurate_2016}.

Unlike the power spectrum, which captures the amplitude of fluctuations as a function of scale, the bispectrum encodes information about the scale-dependent coupling of modes that arise from non-linear structure formation under gravitational instability. The bispectrum, $B$, contains the lowest-order non-Gaussian information contained within a field, that is, for a Gaussian field, $B=0$.

Similarly to the convergence power spectrum, $B_{\kappa}$ is a projection of the three-dimensional matter bispectrum, $B_{\delta}$, weighted by the lensing kernel along the line of sight:

\begin{equation}
    B_{\kappa}(\ell_1,\ell_2,\ell_3) = \int_0^{\chi_*} d\chi \frac{W(\chi)^3}{r(\chi)^4}B_{\delta}(k_1, k_2, k_3; z(\chi)),
\end{equation}
where $W(\chi)$ is the lensing weight function defined in Section~\ref{sec:stats_power_spectrum}, and $k_i = \ell_i/\chi$ are three-dimensional wavenumbers corresponding to the angular modes $\ell_i$. 

We note that for simplicity, we have presented the definitions of the power spectrum and bispectrum in the flat sky approximation. When measuring these statistics from the convergence map, we will not rely on this approximation.

\subsection{Probability Distribution Function (PDF)}\label{sec:stats_PDF}
Unlike the power spectrum or the bispectrum, which focus on spatial correlations, the PDF of the convergence field captures the global properties of the convergence field by describing the distribution of the convergence values at every point in the sky. 

The one-point PDF is defined as the probability that the convergence field will have a particular value at a given point in the sky, or at a particular pixel in the $\kappa$ map. Mathematically, the PDF can be approximated as:
\begin{equation}\label{eq:PDF_hist}
    \mathcal{P}(\kappa) = \frac{N(\kappa)}{\Delta\kappa N_{\mathrm{\rm pix}}},
\end{equation}
where $N(\kappa)$ is the number of pixels within the map that have convergence $\kappa$, $\Delta\kappa$ is the histogram bin width and $N_{\mathrm{\rm pix}}$ is the total number of pixels within the map. The PDF of the convergence field captures the full information on the one-point statistics of density fluctuations projected along the line-of-sight.

MG theories predict distinct signatures on the one-point PDF of the density field due to the altered dynamics of structure formation. In particular, skewness and kurtosis will be different \citep{hellwing_revealing_2017}, and the PDF can help to discriminate between different cosmological scenarios \citep{gough_one-point_2022}.

\subsection{Peak counts}\label{sec:stats_peaks}

Peak counts in the convergence field capture the density of local maxima in the convergence field, which are typically associated with higher-density regions and cosmic structures such as halos or clusters of galaxies. By identifying regions in the convergence maps where $\kappa$ reaches a certain value, we can analyse the number of peaks and their respective heights and spatial distribution. This accounts for the positions and amplitudes of the peaks, and provides insights into the underlying matter distribution and properties of the lensing field, and hence the underlying cosmology.

A peak is defined as a region where the value of $\kappa$ exceeds that of its surrounding neighbours, {\it i.e.} in a 2D map, the value of $\kappa$ within a pixel must be greater than that of its surrounding eight pixels. The total peak count, $N_{{\rm peaks}}$, can be expressed as a function of a threshold $\kappa$ value.

Peak counts capture the occurrence of extreme events, like the formation of massive halos, which are not captured by statistics such as the power spectrum. MG theories often predict enhanced structure formation on large scales, and hence a higher abundance of convergence field peaks compared to GR \citep{higuchi_imprint_2016, liu_constraining_2016, shirasaki_imprint_2017, davies_constraining_2024}. 

\subsection{Minkowski Functionals (MFs)}\label{sec:stats_MFs}

Minkowski functionals (MFs) are a set of shape descriptors that characterise the topological and morphological information contained within a field (see $e.g.$ \citealt{mecke_robust_1993}). For a 2D field like the convergence map, there are three types of MFs: $V_0$, $V_1$ and $V_2$, representing the area within which $\kappa$ is above a threshold value, the total length of boundaries above a threshold, and the integrated curvature along the contours above the threshold, respectively. By computing the three MFs with respect to a given threshold value of $\kappa$, we can quantify how the morphology of the convergence field changes as a function of threshold. 

The three MFs for a 2D field are defined as follows:
\begin{equation}\label{eq:MFs_V0}
    V_0(\kappa) = \int_{\Sigma(\kappa)} dA,
\end{equation}
\begin{equation}\label{eq:MFs_V1}
    V_1(\kappa) = \frac{1}{4}\int_{\partial \Sigma(\kappa)} dl,
\end{equation}
\begin{equation}\label{eq:MFs_V2}
    V_2(\kappa) = \frac{1}{2\pi}\int_{\partial\Sigma(\kappa)} \mathcal{K} dl,
\end{equation}
where $A$ is the area contained within a boundary, $l$ is the path length of the boundary and $\mathcal{K}$ is the geodesic curvature of the boundary. MFs focus on the topological features within a field, providing complementary information to other statistics such as the power spectrum or peak counts. In particular, $V_0$ characterises the overall intensity of the convergence map, {\it i.e.,} it is a measure of how much of the map contains values over a certain threshold. $V_1$ provides insights into the complexity of the topological features of the convergence field. $V_2$ is the Euler characteristic that reflects the connectedness of the field, {\it i.e.} the number of connected components or voids within the map. 

Structure formation under different MG models impacts the topology and morphology of the large-scale structure, leaving signatures in the MFs \citep{ling_distinguishing_2015, shirasaki_imprint_2017, jiang_minkowski_2024}. 

\begin{figure*}
  \centering
  \includegraphics[width=\textwidth]{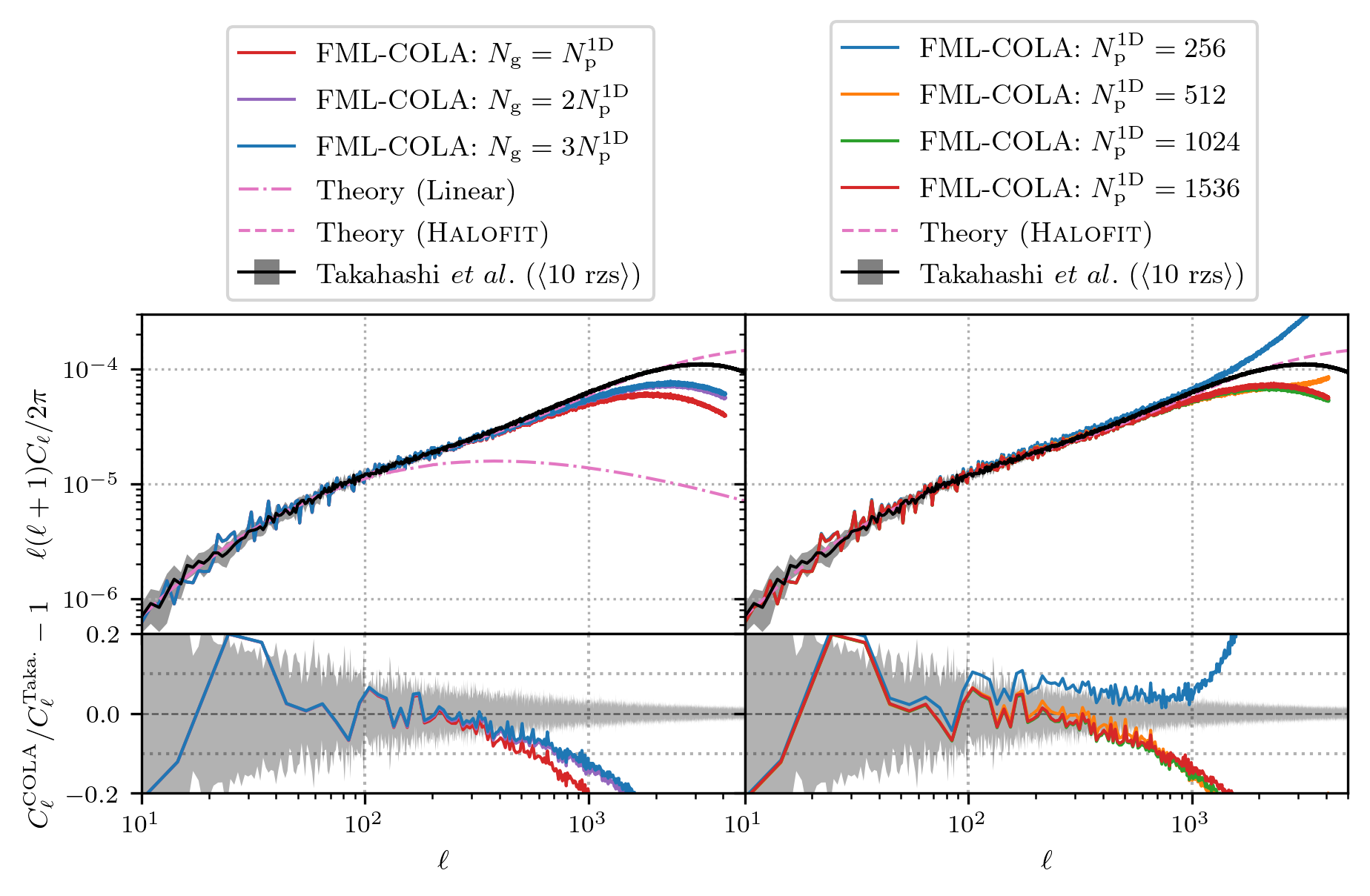}
  \caption{Convergence field power spectrum at $z=0$ computed by \FML{}, {\it N}-body in \taka{} and theoretical predictions. Shaded regions in the top and bottom panels show the standard deviation of $10$ \taka{} realisations. In the left panel, we show the effect of varying the \FML{} force resolution, $N_{\mathrm{g}}$, while keeping the number of particles (and mass resolution) fixed to $N_{\mathrm{p}}^{\mathrm{1D}}=1536$. For $\ell\lesssim750$, we do not observe appreciable improvements to the accuracy beyond $N_{\mathrm{g}}\geq2N_{\mathrm{p}}^{\mathrm{1D}}$. In the right panel, we vary the particle number, $N_{\mathrm{p}}^{\mathrm{1D}}$, whilst the force resolution is fixed to $N_{\mathrm{g}}=2N_{\mathrm{p}}^{\mathrm{1D}}$. We see convergence for $N_{\mathrm{p}}^{\mathrm{1D}}\geq512$ for $\ell\lesssim750$.}
  \label{fig:Cls_Nbody_v_COLA}
\end{figure*}

\begin{figure*}
  \centering
  \includegraphics[width=\textwidth]{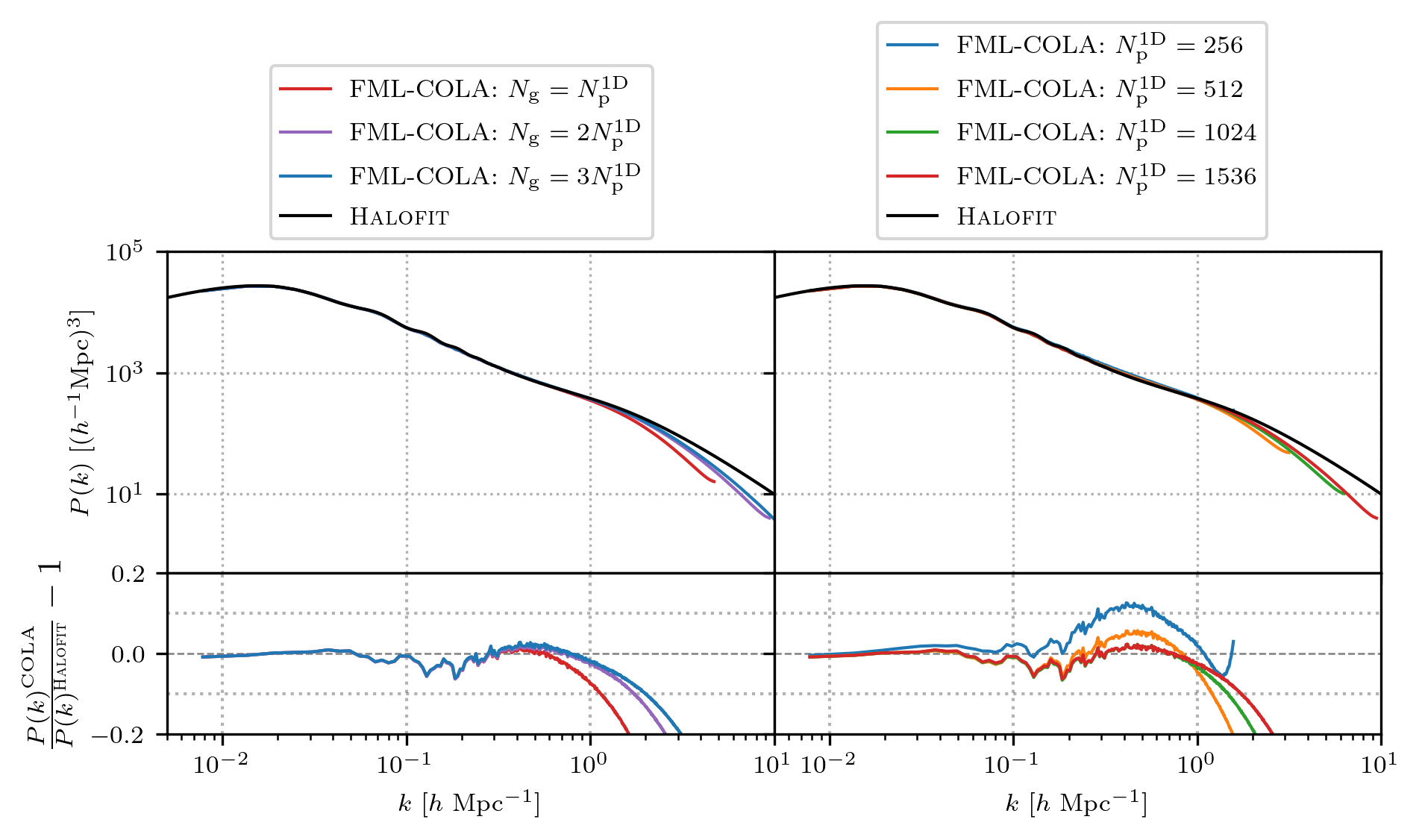}
  \caption{Dark matter power spectrum, $P(k)$, at $z=0$ computed with \FML{} and Halofit. In the left panel, we show the results of varying \FML{} force resolution, $N_{\mathrm{g}}$, while keeping the number of particles fixed to $N_{\mathrm{p}}^{\mathrm{1D}}=1536$. Convergence is reached for $N_{\mathrm{g}}\geq2N_{\mathrm{p}}^{\mathrm{1D}}$ at scales $k\sim1 h$ Mpc$^{-1}$. In the right panel, we vary the particle number, $N_{\mathrm{p}}^{\mathrm{1D}}$, while the force resolution is fixed to $N_{\mathrm{g}}=2N_{\mathrm{p}}^{\mathrm{1D}}$. For $k\sim0.5 h$ Mpc$^{-1}$, we have convergence at $N_{\mathrm{p}}^{\mathrm{1D}}\geq512$.  }
  \label{fig:Pofk_Nbody_v_COLA}
\end{figure*}

\section{Validation in GR}\label{sec:validation}

In this section, we demonstrate the accuracy of the weak lensing convergence maps obtained with the new \FML{} method by examining their performance on the five summary statistics defined in Section~\ref{sec:stats}. In particular, we validate our lightcone implementation by comparing the statistics computed from our maps against those computed on maps generated with high-resolution, full {\it N}-body simulations presented in \cite{takahashi_full-sky_2017}. In the following, we will refer to these maps \taka{}. 

The remainder of this section is structured as follows: in Section~\ref{sec:validation_sim_specs}, we present the specifications of the \FML{} and \taka{} simulations used to obtain the $z=0$ convergence maps, and a comparison of the power spectrum, bispectrum, PDF, peak counts and MFs of the two sets of maps are presented in Sections \ref{sec:validation_power_spectrum}--\ref{sec:validation_MFs}, respectively. 

\subsection{Simulation specifications}\label{sec:validation_sim_specs}
The cosmology chosen in this work is consistent with the WMAP Year 9 results \citep{hinshaw_nine-year_2013} to match the \taka{}: dark matter density parameter $\Omega_{\rm CDM}=0.233$, baryon density $\Omega_b=0.046$, total matter density $\Omega_m=\Omega_{\rm CDM}+\Omega_b=0.279$, cosmological constant $\Omega_{\Lambda}=0.721$, Hubble parameter $h=0.7$, amplitude of density fluctuations $\sigma_8=0.82$ and the scalar spectral index $n_s=0.97$. 

The \taka{} simulations have the following set-up: the initial conditions are generated using 2LPT, and $N_{\mathrm{p}}^{\mathrm{1D}}=2048$ dark matter particles are evolved in 14 cubic boxes of increasing volume, where the side length of the simulation box varies from $L_{\mathrm{box}}=450\ h^{-1}$Mpc to $L_{\mathrm{box}}=6300\ h^{-1}$Mpc. This leads to smaller boxes having a higher mass resolution. Within each box, three spherical lens shells are constructed, and the public code \textsc{GrayTrix} \citep{shirasaki_probing_2015} is used to project the positions of the particles onto these shells. Three pixel resolutions are adopted in \taka{}, and we compare our work to the maps having $N_{\mathrm{side}}=4096$. We take the results corresponding to a source redshift of $z_* = 1.0334$. We compute each statistic for $10$ realisations of the full sky map to estimate errors. The smallest Fourier scale that can be observed, $\ell_{\rm max}$, is approximately $\ell_{\rm max}\sim2N_{\rm side}$. 

All \FML{} runs presented in this work use the same realisation of Gaussian initial conditions, where the Fourier amplitudes at the initial redshift, $z_{\rm ini}$, have been rescaled from the amplitudes of the linear matter power spectrum at $z=0$, obtained using the Boltzmann solver \textsc{CAMB}\footnote{\url{https://camb.readthedocs.io/en/latest/}} \citep{lewis_efficient_2000} for the chosen cosmology. These initial conditions are assigned using 2LPT, and are evolved from $z_{\rm ini}=20$ to the present day $z=0$ in 40 timesteps, in a cubic box of side length $L_{\mathrm{box}}=1024$ $h^{-1}$Mpc. To reduce the effects of cosmic variance, we run pair-fixed simulations \citep{pontzen_inverted_2016}.

For all the statistics considered, we investigate how the force resolution of \FML{} affects the agreement with the \taka{} results. The force resolution is determined by the number of grid cells along each length of the simulated cubic volume, $N_{\mathrm{g}}$, within which the Poisson equation is solved in Fourier space to obtain the gravitational potential. $N_{\mathrm{g}}$ is usually chosen according to the number of DM particles along one dimension within the simulation, $N_{\mathrm{p}}^{\mathrm{1D}}$. We choose $N_{\mathrm{g}}=1,2,3 \times N_{\mathrm{p}}^{\mathrm{1D}}$, where $N_{\mathrm{p}}^{\mathrm{1D}}=1536$. This corresponds to a physical resolution of 0.67, 0.33 and 0.22 $h^{-1}$Mpc, respectively.

We also investigate how the mass resolution of the \FML{} simulation impacts the accuracy of the maps: for this, we compare simulations run with $N_{\mathrm{p}}^{\mathrm{1D}}=256, 512, 1024, 1536$, where the force resolution is fixed at $N_{\mathrm{g}}=2N_{\mathrm{p}}^{\mathrm{1D}}$. The corresponding physical resolutions for these specifications are 2.00, 1.00, 0.50 and 0.33 $h^{-1}$Mpc, respectively. Note that we only perform this test for the convergence field power spectrum, where we see that $N_{\mathrm{p}}^{\mathrm{1D}}=1536$ achieves more accurate $C_{\ell}^{\kappa\kappa}$s at $\ell \sim 1000$. Therefore, we set $N_{\mathrm{p}}^{\mathrm{1D}}=1536$ for all statistics subsequently considered.

In all \FML{} simulations, the lightcone switches on at $z_*=1.0334$ to match the chosen \taka{} outputs. To cover the entire sky ($f_{\rm sky}=1$), 8 box replicas (2 replicas for each dimension) are used, resulting in an effective box size of $L_{\rm box} = 2048 h^{-1}$ Mpc. The concentric shells from $z_*$ to the observer at $z=0$ are binned in intervals of the scale factor, as described in Section~\ref{sec:WL_lightcone}, with the binning width set to $\Delta a=0.025$. All statistics presented in this work are computed using the convergence field at $z=0$. Finally, the \textsc{HEALPix} convergence maps constructed on-the-fly at each interval of $\Delta a$ are computed with a pixel resolution of $N_{\rm side}^{\rm COLA}=2048$.

\begin{figure*}
  \centering
    \includegraphics[width=\textwidth]{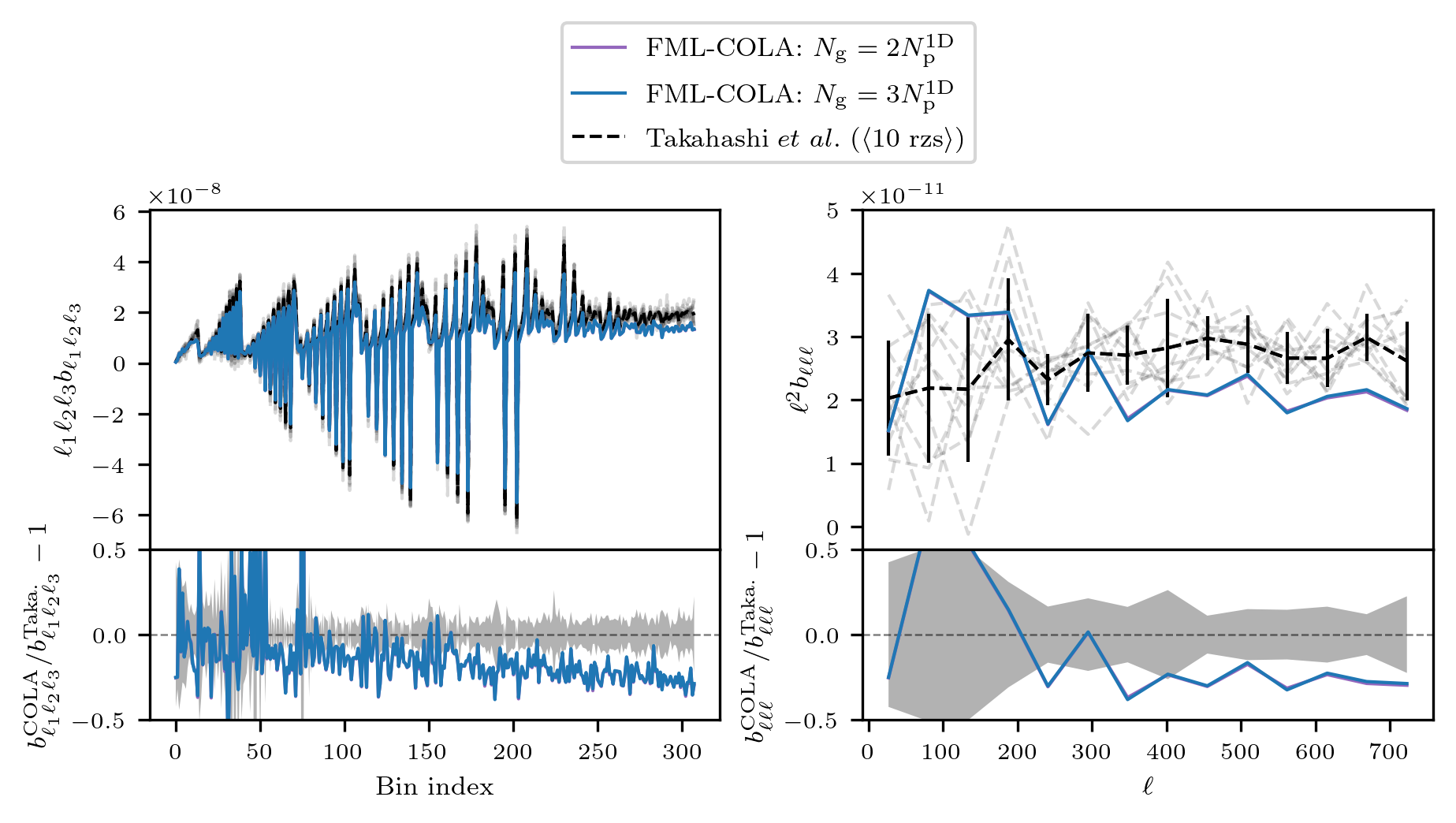}
  \caption{Convergence field bispectrum at $z=0$, as computed by \FML{} and {\it N}-body in \taka{}. Shaded regions correspond to errors obtained from $10$ \taka{} realisations. The bispectrum is computed using a total number of bins $N_{\mathrm{bins}}=15$ within $10<\ell<750$. In the left panel, we show all ($\ell_1,\ell_2,\ell_3$) configurations satisfying the triangle condition, whereas in the right panel, we show equilateral configurations only ($\ell \equiv \ell_1=\ell_2=\ell_3$). \FML{} underestimates the bispectrum for increasing bin index given the limitations of approximate methods. }
  \label{fig:bispectrum_Nbody_v_COLA}
\end{figure*}

\begin{table}
\caption{In cases where the convergence maps' angular pixel resolution, $N_{\rm side}$, is downgraded ({\it i.e.} when considering the PDF, peak counts and MFs in this work), Gaussian smoothing is applied to reduce the effects of pixelation, and the maps are normalised by their standard deviation, $\sigma_{\rm pix}$. For every $N_{\rm side}$ considered in this work, we show the pixel size in arcmins, $\theta_{\rm pix}$, the chosen smoothing scale, $\theta_G=1.45\theta_{\rm pix}$, and the standard deviation of the pixels in the \textsc{FML-COLA} maps and {\it N}-body maps, $\sigma_{\rm pix}^{\rm COLA}$ and $\sigma_{\rm pix}^{{\rm {\it N}\text{-}body}}$, respectively. We also show the angular resolution corresponding to the $N_{\rm side}$ at which the $N$-body maps were generated. The values of $\sigma_{\rm pix}^{{\rm COLA}}$ are computed from simulations with force resolution $N_{\mathrm{g}}=2N_{\mathrm{p}}^{\mathrm{1D}}$. We see better agreement between the \textsc{FML-COLA} and \taka{} map standard deviations at lower resolution. }
\centering
 \begin{tabular}{||c|c|c|c|c||} 
 \hline
 $N_{\rm side}$ & $\theta_{\rm pix}$ (arcmins) & $\theta_G$  (arcmins) & $\sigma_{\rm pix}^{\rm COLA}$ & $\sigma_{\rm pix}^{{\rm {\it N}\text{-}body}}$ \\ [0.5ex] 
 \hline\hline
 128 & 27.48 & 39.85 & 0.00418 & 0.00416 \\ 
 256 & 13.74 & 19.93 & 0.00559 & 0.00561 \\
 512 & 6.87 & 9.96 & 0.00729 & 0.00741 \\
 1024 & 3.44 & 4.98 & 0.00933 & 0.00970\\
 2048 & 1.71 & 2.49 & 0.01154 & 0.01235 \\ 
 4096 &  0.86 & ---  &  ---  &  ---  \\ [1ex] 
 \hline
 \end{tabular}
 \label{tab:smoothing_scale}
\end{table}

When considering the power spectrum and the bispectrum, the $\Tilde{\kappa}$-maps obtained with \FML{} and \taka{} are kept at the resolution at which they were generated (recall that the maximum scale captured by a given $N_{\rm side}$ is $N_{\rm side}\sim 2 \ell_{\rm max}$). However, for the PDF, peak counts and MFs, we investigate how the accuracy of the observed convergence field properties depend on the map's resolution. In particular, both the \FML{} and \taka{} maps are downgraded to $N_{\rm side}=128,256,512,1024,2048$, where the value of a pixel in the downgraded map is set to the mean of the corresponding pixels in the higher-resolution map before downgrading. To remove the effects of pixelization after downgrading, we apply Gaussian smoothing with a kernel width, $\theta_G$, determined by the pixel size, $\theta_{\rm pix}$, such that $\theta_G = 1.45\theta_{\rm pix}$. Table ~\ref{tab:smoothing_scale} shows the corresponding value of $\theta_{\rm pix}$ and $\theta_G$ for each value of $N_{\mathrm{side}}$ we consider. 

The \FML{} and \taka{} maps are then normalised by their pixel standard deviation, which we denote $\sigma_{\rm pix}^{\rm COLA}$ and $\sigma_{\rm pix}^{{\rm {\it N}\text{-}body}}$, respectively (also shown in Table \ref{tab:smoothing_scale}). Normalising the maps in this way reduces the effects of noise and allows for better comparison between the \FML{} and \taka{} results, as the maps are brought to a common scale where we can better identify the shape of the distributions, as opposed to the overall amplitudes that may vary with resolution. Table \ref{tab:smoothing_scale} shows that the standard deviation between the \textsc{FML-COLA} and \taka{} maps is more similar for lower resolution. The downgraded, smoothed and normalised convergence maps are denoted $\Tilde{\kappa}$ throughout the remainder of this paper. 

For the convergence tests on pixel resolution, we set the \FML{} force resolution to $N_{\mathrm{g}}=2N_{\mathrm{p}}^{\mathrm{1D}}$. The $N_{\mathrm{g}}$ convergence tests on the PDF, peak counts and MFs are then conducted with the value of $N_{\mathrm{side}}$ that achieves the best accuracy as compared to the \taka{} maps ($N_{\mathrm{side}}=256$ for the PDF and peak counts, and $N_{\mathrm{side}}=128$ for the MFs).

\subsection{Power spectrum}\label{sec:validation_power_spectrum}

\begin{figure*} 
  \centering
  \includegraphics[width=\textwidth]{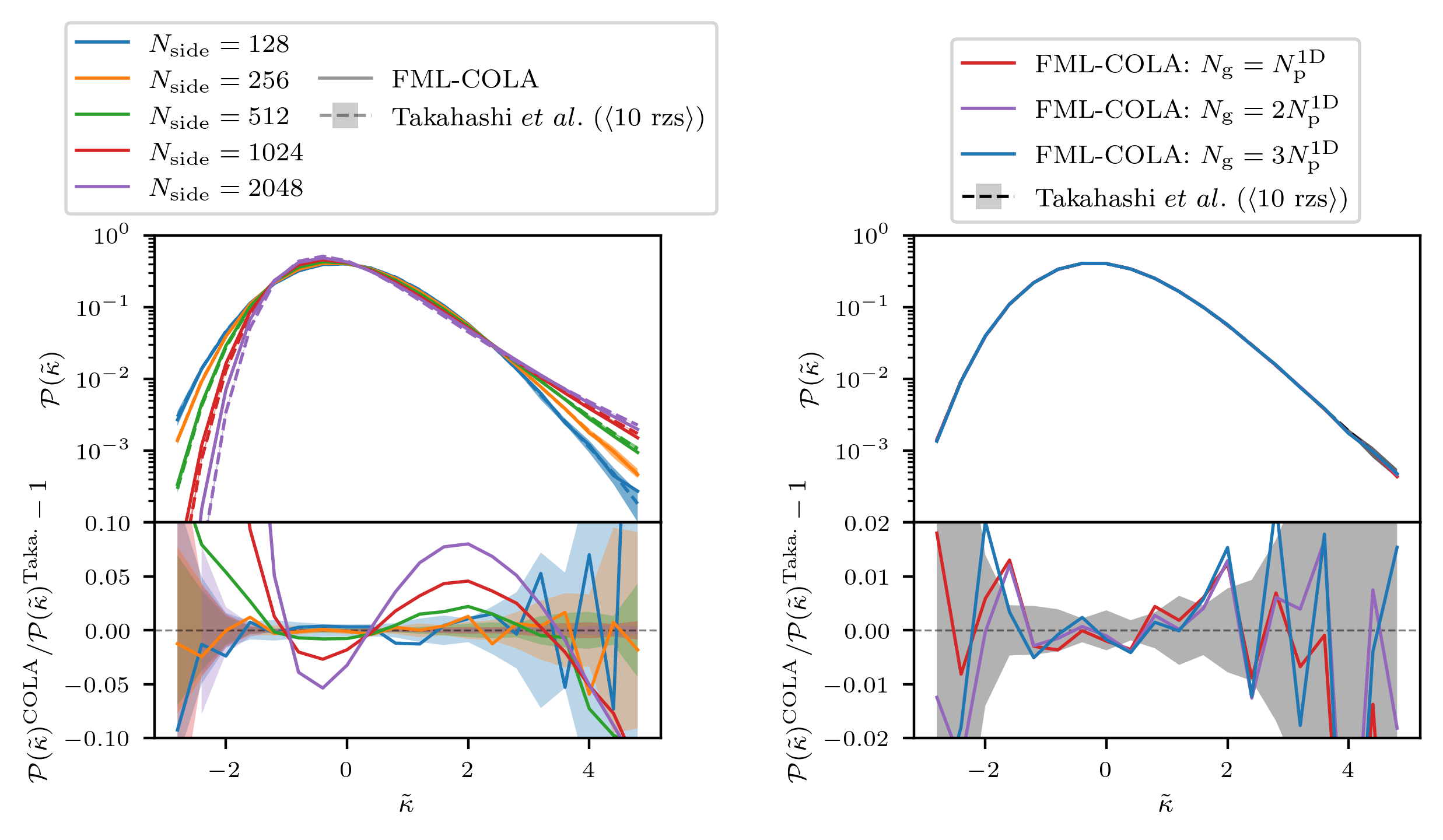}
  \caption{Pixel PDF of the convergence fields computed with \FML{} and {\it N}-body simulations in \taka{}. Shaded regions show the standard deviation of $10$ \taka{} realisations. Each map is smoothed with a Gaussian kernel according to the pixel size, and the smoothing scale for each value of $N_{\rm side}$ is reported in Table \ref{tab:smoothing_scale}. In the left panel, we show the effect of downgrading \FML{} and \taka{} maps to the same $N_{\rm side}$, for values $N_{\rm side}=128,256,1024,2048$. For this, we set the \FML{} force resolution to $N_{\mathrm{g}}=2N_{\mathrm{p}}^{\mathrm{1D}}$. The PDF of \FML{} maps shows better agreement with that of \taka{} for $N_{\rm side}\leq256$. In the right panel, we show the impact of varying the \FML{} force resolution, $N_{\mathrm{g}}$, after downgrading the \FML{} and \taka{} maps to $N_{\rm side}=256$. The choice of $N_{\mathrm{g}}$ does not make a significant difference in P($\Tilde{\kappa}$).  }
  \label{fig:PDF_Nbody_v_COLA}
\end{figure*}

Figure~\ref{fig:Cls_Nbody_v_COLA} shows the comparison between the convergence power spectrum, $C_{\ell}^{\kappa\kappa}$, computed from the \FML{} maps and that computed from the maps obtained from \taka{}. For the \FML{} prediction, $C_{\ell}^{\kappa\kappa}$ is computed on-the-fly following the method described in Section~\ref{sec:WL_map_construction}. For the \taka{} maps, $C_{\ell}^{\kappa\kappa}$ is computed using \textsc{HEALPix}. In Figure ~\ref{fig:Cls_Nbody_v_COLA}, we also show the predictions from linear and non-linear (\textsc{Halofit}) theory, which are computed by integrating the matter power spectra, $P(k,z)$, obtained with \textsc{CAMB}\footnote{CAMB uses \textsc{Halofit} to compute the non-linear power spectrum. We use the implementation of \citealt{mead_accurate_2015} provided in version 1.0.8 \citep{antony_lewis_2019_3452064}.} \citep{antony_lewis_2019_3452064} from $z_*$ to $z=0$ according to Eq.~\ref{eq:C(ell)}, where the comoving distance to redshift $z$ is defined as:

\begin{equation}
    \chi(z) = \int_0^{z} \frac{d_H}{E(z')}\ dz', \\
\end{equation}
where $d_H = c/H_0$ is the Hubble distance, and $E(z)$ is the dimensionless Hubble function:
\begin{equation}
    E(z) = [\Omega_M(1\ +\ z)^3 + \Omega_{\Lambda}]^{1/2},
\end{equation}
where $\Omega_m$ and $\Omega_{\Lambda}$ are the matter and cosmological constant density parameters, respectively.

Figure ~\ref{fig:Cls_Nbody_v_COLA} shows that \FML{} accurately captures the power on linear scales, over the range $\ell\sim200$ where linear theory is valid. It also agrees well with the non-linear prediction up to $\ell\sim700$, but underestimates the power on smaller scales given the limited resolution of the Particle-Mesh method and the approximation of using LPT.

\begin{figure*} 
  \centering
  \includegraphics[width=\textwidth]{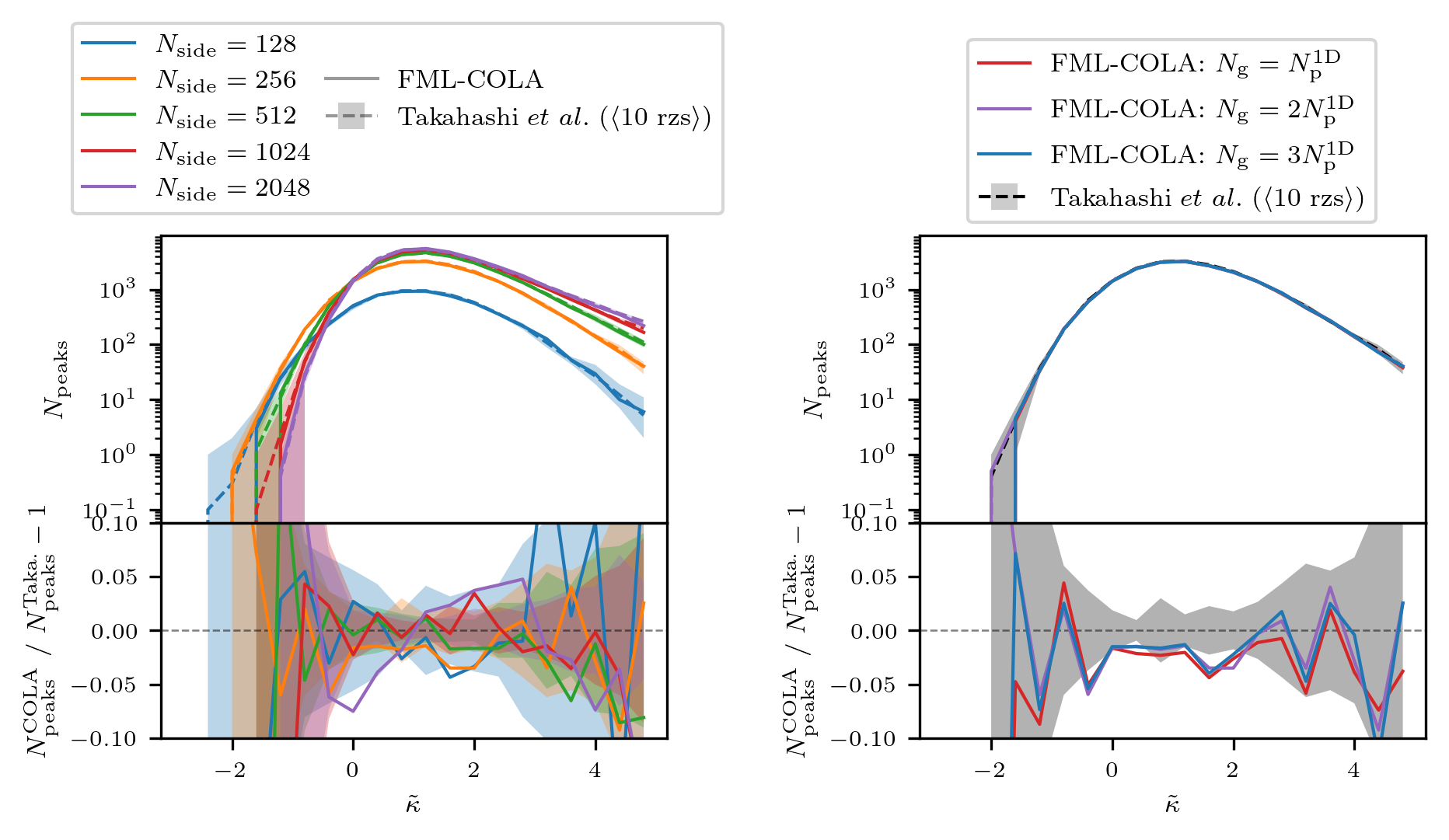}
  \caption{Peak counts, $N_{\rm peaks}$, of the $z=0$ convergence fields, $\Tilde{\kappa}$, computed with \FML{} and {\it N}-body simulations in \taka{}. Shaded regions show the standard deviation of $10$ \taka{} realisations. Gaussian smoothing is applied to the convergence maps, where the smoothing scale is given in Table \ref{tab:smoothing_scale}. In the left panel, we show the impact of varying the pixel resolution, $N_{\rm side}$, by downgrading both \FML{} and \taka{} maps to $N_{\rm side}=128,256,1024,2048$. The \FML{} simulations are run with force resolution $N_{\mathrm{g}}=2N_{\mathrm{p}}^{\mathrm{1D}}$. For all pixel resolutions, the \FML{} convergence fields do well in recovering the \taka{} peak counts. In the right panel, we vary the \FML{} force resolution, $N_{\mathrm{g}}$, where all maps are downgraded to $N_{\rm side}=256$. The peak count is fairly insensitive to force resolution at this map resolution. }
  \label{fig:peaks_Nbody_v_COLA_Nside_Nmesh}
\end{figure*}

We first investigate how the force resolution of the \FML{} simulations, $N_{\mathrm{g}}$, affects the accuracy of the predicted power spectrum. The left plot of Figure~\ref{fig:Cls_Nbody_v_COLA} shows the results from \FML{} simulations run with $N_{\mathrm{g}}=1,2,3\times N_{\mathrm{p}}^{\mathrm{1D}}$, where $N_{\mathrm{p}}^{\mathrm{1D}}=1536$. A lower force resolution shows poorer agreement with the {\it N}-body results ($\sim75\%$ less agreement at $\ell\sim1000$ between $N_{\mathrm{g}} = 1N_{\mathrm{p}}^{\mathrm{1D}}$ and $N_{\mathrm{g}} = 2N_{\mathrm{p}}^{\mathrm{1D}}$), but we do not see a significant improvement in the accuracy of $C^{\kappa\kappa}_{\ell}$ when increasing $N_{\mathrm{g}}$ up to $N_{\mathrm{g}} = 3N_{\mathrm{p}}^{\mathrm{1D}}$, for the same number of timesteps. Using a force resolution of $N_{\mathrm{g}}=2N_{\mathrm{p}}^{\mathrm{1D}}$, the \FML{} method agrees with the \taka{} results to within $\sim5\%$ up to $\ell\sim500$, and to within $\sim10\%$ up to $\ell\sim750$.

Next, we compare the power spectrum computed by \FML{} for different mass resolutions, as shown in the right plot of Figure \ref{fig:Cls_Nbody_v_COLA}. For this test, we vary $N_{\mathrm{p}}^{\mathrm{1D}}$ in $N_{\mathrm{p}}^{\mathrm{1D}}=256,512,1024,1536$, after fixing $N_{\mathrm{g}}=2N_{\mathrm{p}}^{\mathrm{1D}}$. For $N_{\mathrm{p}}^{\mathrm{1D}}=256$, the mass resolution is too low and the agreement to \taka{} at $\ell\sim200$ is less than $5\%$. Figure~\ref{fig:Cls_Nbody_v_COLA} shows an excess in power at smaller scales at this mass resolution. This arises from particle shot noise which scales with the particle number density, $n$, as $1/n$. When increasing the mass resolution beyond, $N_{\mathrm{p}}^{\mathrm{1D}}\geq512$, \FML{} agrees with \taka{} very well. Using a higher mass resolution improves our ability to resolve smaller scales, which can be seen at $\ell\sim1000$. 

Comparisons of the angular power spectrum of the weak lensing convergence field predicted by \FML{} and \taka{} show that the choice of $N_{\mathrm{p}}^{\mathrm{1D}}=1536$, $N_{\mathrm{g}}=2N_{\mathrm{p}}^{\mathrm{1D}}$ is adequate to predict the power spectrum to within $5\%$ up to $\ell \sim 500$, and within $10\%$ up to $\ell \sim 750$. In Figure ~\ref{fig:Pofk_Nbody_v_COLA}, we show results of the $N_{\mathrm{g}}$ and $N_{\mathrm{p}}^{\text{1D}}$ convergence tests on the dark matter power spectrum, $P(k)$, for comparison. The \FML{} results are again computed on-the-fly during the simulation, and we compare the results of \textsc{Halofit} $P(k)$ at $z=0.0$ obtained with \textsc{CAMB}. This comparison of the matter $P(k)$ further justifies our choice of $N_{\rm p}^{\rm 1D}=1536$ in all subsequent results. To further improve the accuracy of the \FML{} maps, the number of timesteps should be increased \citep{fiorini_fast_2021} at the expense of the speed of simulations.


\subsection{Bispectrum}\label{sec:validation_bispectrum}

Given that the angular power spectrum of the convergence field, computed using \FML{}, agrees with the \taka{} results to within $10\%$ for $\ell\lesssim750$, we compute the bispectrum up to $\ell_{\rm max}=750$. The bispectrum is computed using \textsc{PolySpec}\footnote{\url{https://github.com/oliverphilcox/PolySpec}.} \citep{philcox_optimal_2023} for a total number of bins $N_{\mathrm{bins}}=15$ within $10<\ell<750$. Increasing the bin index corresponds to increasing $\ell_3$ first, then $\ell_2$, then $\ell_1$, taking all $\ell_1,\ell_2,\ell_3$ configurations within the $\ell$ range of each bin that satisfy the triangle condition ($\ell_1 \leq \ell_2 + \ell_3$ for $\ell_1 \geq \ell_2, \ell_3$). For all configurations, a bin index of 150 corresponds to $\ell_1,\ell_2,\ell_3=241, 723, 723$, respectively, which corresponds to the loss of power in the \FML{} $C_{\ell}^{\kappa\kappa}$ for $\ell\gtrsim750$. A check was made to ensure that \textsc{PolySpec} gives the same $C_{\ell}^{\kappa\kappa}$ prediction from the \FML{}-generated $\Tilde{\kappa}$-map as is computed on-the-fly with \FML{}. 

Figure~\ref{fig:bispectrum_Nbody_v_COLA} shows a comparison of the bispectrum computed from weak lensing maps obtained from \FML{} and \taka{} simulations, where the force resolution of the \FML{} simulations is $N_{\mathrm{g}}=2,3 \times N_{\mathrm{p}}^{\mathrm{1D}}$, for $N_{\mathrm{p}}^{\mathrm{1D}}=1536$. The left plot shows the bispectrum when considering all triangle configurations, whereas the right plot shows equilateral configurations ($\ell \equiv \ell_1=\ell_2=\ell_3$), plotted with respect to $\ell$. 

\FML{} underestimates the bispectrum by $\sim 25\%$ for increasing $\ell$ compared to the \taka{} results, for both $N_{\mathrm{g}}=2,3 \times N_{\mathrm{p}}^{\mathrm{1D}}$, due to the limitations of approximate methods. The force resolution and number of timesteps are relatively small in these \FML{} runs; increasing these parameters would increase the accuracy, however, this is the trade-off between speed and accuracy that we expect when using approximate methods. 

\subsection{PDF}\label{sec:validation_PDF}

The next statistic we consider is the PDF of the convergence field, where the convergence maps, $\Tilde{\kappa}$, are generated following the method described in Section~\ref{sec:validation_sim_specs}. The PDF is computed using a histogram in 21 bins from $-3 <\ \Tilde{\kappa}\ <5$ following Eq.~\ref{eq:PDF_hist}, where the total number of pixels is $N_{\mathrm{\rm pix}}=12N_{\mathrm{side}}^2=786,432$ and the binning width is $\Delta \Tilde{\kappa} \sim0.38$.

We first test how the angular resolution of the convergence maps generated by \FML{} and \taka{} affects the agreement between the predicted $P(\Tilde{\kappa})$. The left panel of Figure \ref{fig:PDF_Nbody_v_COLA} shows that the PDF computed from the maps obtained with \FML{} is within the error bars of the 10 \taka{} realisations for a pixel resolution of $N_{\rm side}\leq256$ across all values of $\Tilde{\kappa}$. We see that for lower $N_{\rm side}$, a more Gaussian distribution is recovered, while for higher resolution, the PDF detects more non-Gaussian features. To consider the PDF of the convergence fields predicted by \FML{}, we must downgrade the maps to a resolution of $N_{\rm side}=256$ to ensure high enough accuracy.  

Next, we investigate whether increasing the force resolution of the \FML{} simulation gives a prediction of $P(\Tilde{\kappa})$ that agrees better with {\it N}-body. Given the agreement between \FML{} and \taka{} for a pixel resolution $N_{\rm side}\leq256$, we downgrade both the \FML{}- and the {\it N}-body-generated $\Tilde{\kappa}$-maps to $N_{\rm side}=256$ for this test. Figure \ref{fig:PDF_Nbody_v_COLA} shows no notable differences from increasing the force resolution of the simulation when considering the PDF of $\Tilde{\kappa}$, given the lower pixel resolution of these maps.
\begin{figure*} 
  \centering
  \includegraphics[width=\textwidth]{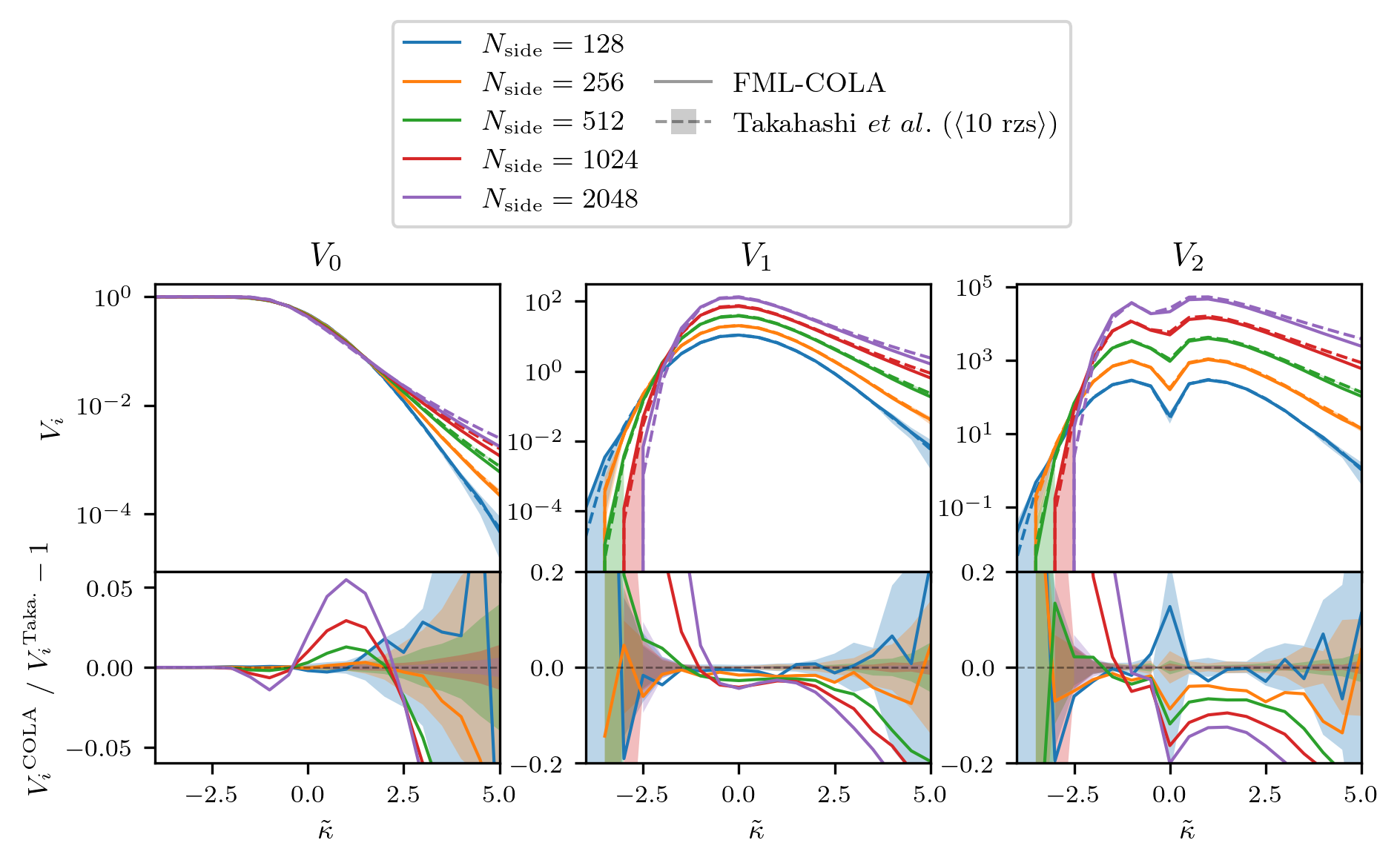}
  \caption{Minkowski functionals of the $z=0$ convergence fields, $\Tilde{\kappa}$, computed with \FML{} (dashed lines) and {\it N}-body in \taka{} (solid lines): $V_0$ (left panel); $V_1$ (middle panel); $V_2$ (right panel). Different colours correspond to the pixel resolution $N_{\rm side}$ of the \FML{} and \taka{} maps. All maps have been smoothed according to the pixel size, where the corresponding smoothing scale for each value of $N_{\rm side}$ is shown in Table \ref{tab:smoothing_scale}. Shaded regions show the standard deviation computed from $10$ \taka{} realisations. For all Minkowski functionals, a pixel resolution of $N_{\rm side}=128$ gives the required accuracy of the \FML{} predictions. 
  }
  \label{fig:MFs_Nbody_v_COLA_Nside}
\end{figure*}

\subsection{Peak counts}\label{sec:validation_peaks}

The peak counts of the convergence field, $\Tilde{\kappa}$, are computed using \textsc{LensTools}\footnote{\url{https://lenstools.readthedocs.io/en/latest/index.html}} \citep{petri_mocking_2016}, where peaks are defined as pixels with a convergence value higher than its 8 surrounding neighbours. We compute the number of peaks for a given value of $\Tilde{\kappa}$, $N_{\rm peaks}$, using $21$ bins over the range $-3 <\ \Tilde{\kappa}\ < 5$.

Figure~\ref{fig:peaks_Nbody_v_COLA_Nside_Nmesh} shows a convergence test for the peak counts of the \FML{} and \taka{} maps when varying the pixel resolution. For all resolutions chosen, the peak counts are reasonably within the respective error bars of the 10 \taka{} realisations. This suggests peaks do not significantly depend on the pixel size within the maps, and are still detectable for lower resolution.  

Next, we present the results of the \FML{} force resolution convergence test in the right panel of Figure~\ref{fig:peaks_Nbody_v_COLA_Nside_Nmesh}. Despite the lack of dependence of $N_{\rm peaks}$ on $N_{\mathrm{side}}$, we set $N_{\mathrm{side}}=256$ for this test as this is also the resolution that better captures the PDF (Section~\ref{sec:validation_PDF}. A higher force resolution of $N_{\mathrm{g}}=2,3 \times N_{\mathrm{p}}^{\mathrm{1D}}$ more accurately recovers peak counts for $-1 \lesssim \Tilde{\kappa} \lesssim 4$, but this improvement in accuracy is marginal, given the lower pixel resolution of the maps.

\begin{figure*} 
  \centering
  \includegraphics[width=\textwidth]{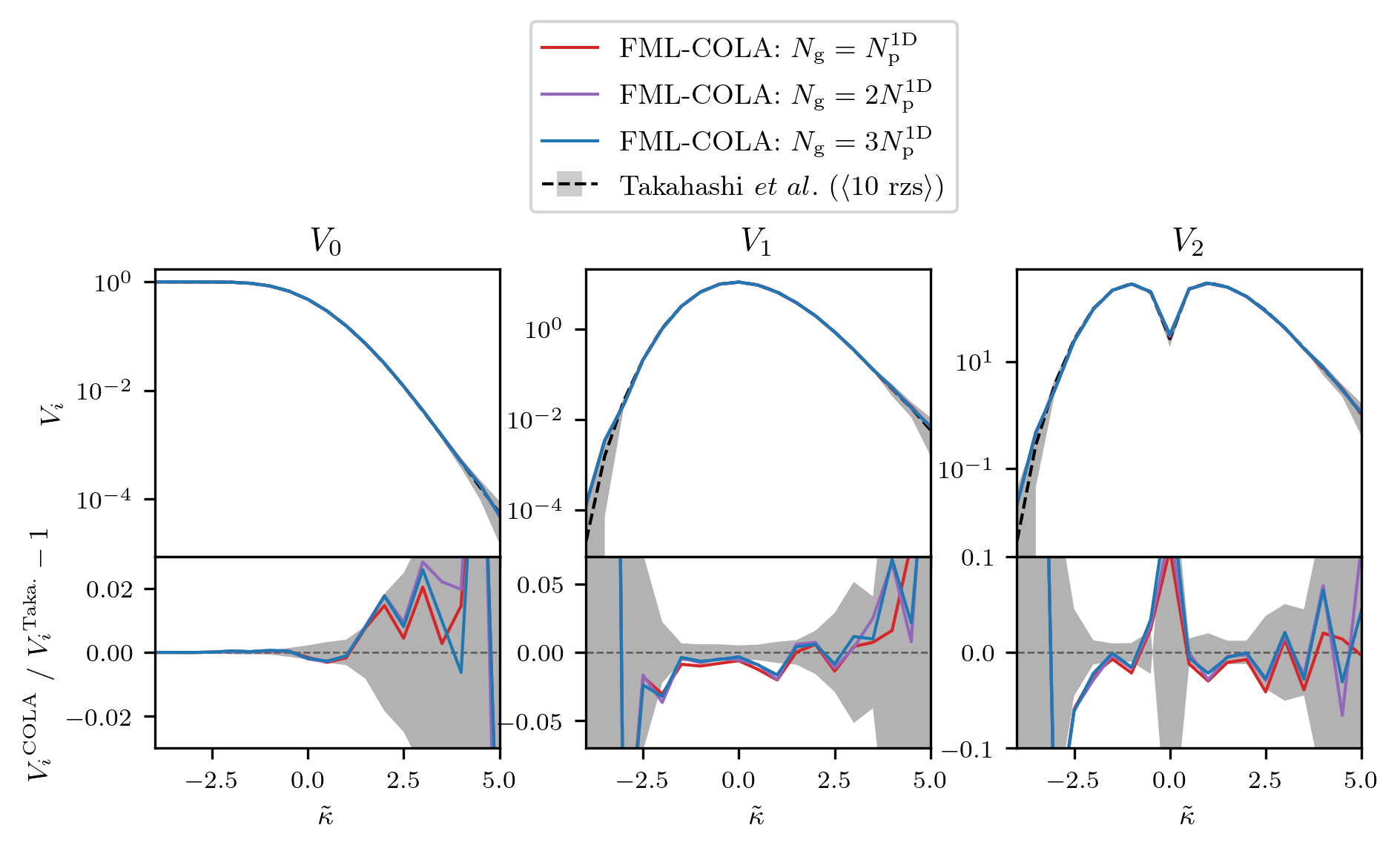}
  \caption{Minkowski functionals of the $z=0$ convergence fields, $\Tilde{\kappa}$, computed with \FML{} (dashed lines) and {\it N}-body in \taka{} (solid lines): $V_0$ (left panel); $V_1$ (middle panel); $V_2$ (right panel). All maps have been downgraded to a pixel resolution $N_{\rm side}=128$ following the results in Figure \ref{fig:MFs_Nbody_v_COLA_Nside}, and have been smoothed by a Gaussian kernel of width 39.85 arcmin, as shown in Table~\ref{tab:smoothing_scale}. Shaded regions correspond to standard deviation of $10$ \taka{} realisations. The \FML{} simulations are run with force resolutions $N_{\mathrm{g}}=1,2,3 \times N_{\mathrm{p}}^{\mathrm{1D}}$ (blue, green and red lines, respectively). There is no significant improvement in the accuracy of the convergence field Minkowski functionals predicted by \FML{} when increasing the simulation force resolution when downgrading the maps to a resolution $N_{\rm side}=128$.}
  \label{fig:MFs_Nbody_v_COLA_Nmesh}
\end{figure*}

\subsection{Minkowski Functionals}\label{sec:validation_MFs}

We use \textsc{Pynkowski}\footnote{\url{https://github.com/javicarron/pynkowski}} \citep{carones_minkowski_2023} to estimate the MFs of the convergence field, $\Tilde{\kappa}$, obtained from \textsc{FML-COLA} and \taka{}. \textsc{Pynkowski} computes $V_0$, $V_1$ and $V_2$ according to Eqs. \ref{eq:MFs_V0}--\ref{eq:MFs_V2}. The MFs are computed in $21$ bins over the range $-5<\Tilde{\kappa}<5$.

Figure~\ref{fig:MFs_Nbody_v_COLA_Nside} shows the results of the pixel resolution convergence test, where we see significant improvements in the \FML{} method in recovering all MFs when downgrading the maps to $N_{\rm side}=128$. Using higher-resolution maps, we see that \FML{} does not recover the same morphological features as \taka{}. To consider the MFs of the convergence fields predicted by \FML{}, we must downgrade the maps to a low resolution of $N_{\rm side}=128$.  

Setting $N_{\rm side}=128$, we then consider the effect of increasing the \FML{} force resolution of each of the three MFs. Figure \ref{fig:MFs_Nbody_v_COLA_Nmesh} shows that, with such a low pixel resolution, the force resolution of \FML{} does not have any significant effect on the accuracy of the convergence field $V_0$, $V_1$ or $V_2$. This force resolution convergence test was also conducted for a slightly higher pixel resolution of $N_{\rm side}=256$, where we see that $N_{\mathrm{g}}=2,3 \times N_{\mathrm{p}}^{\mathrm{1D}}$ does better at recovering the convergence field MFs than $N_{\mathrm{g}}=1N_{\mathrm{p}}^{\mathrm{1D}}$, however, we cannot trust the MFs of \FML{} maps with resolution higher than $N_{\rm side}=128$.

\section{Extension to MG}\label{sec:extension_to_MG}

With \FML{}, we can model structure formation under many theories of MG, which allows for a self-consistent comparison to the predictions of GR. In this section, we extend our analysis of the weak lensing convergence field predicted by \FML{} to two theories: $f(R)$ gravity, where $f_{R0}=10^{-5}$, and nDGP gravity in which $r_c = 1.2 H_0^{-1}$, which will hereafter be referred to as F5 and N1.2, respectively. 

As in Section~\ref{sec:validation} for the validation in GR, we estimate the uncertainty of each statistic using $10$ realisations of the full sky maps of \taka{} in GR. These uncertainties are only indicative, given the small number of realisations used. However, they allow us to compare the ability of each statistic to distinguish between GR and MG.

In Section~\ref{sec:extension_to_MG_sim_specs}, we detail the \FML{} specifications used to obtain the F5 and N1.2 $z=0$ convergence fields, and analyses of the same five statistics are provided in Sections \ref{sec:extension_to_MG_power_spectrum}--\ref{sec:extension_to_MG_MFs}, respectively. In particular, we focus on the ratio between the predictions of MG and GR, to examine which statistics could likely enable MG detection. Finally, in Section~\ref{sec:extension_to_MG_resources}, we present the computational resources required for the GR, F5 and N1.2 \FML{} runs.

\subsection{Simulation specifications}\label{sec:extension_to_MG_sim_specs}

A total of three \FML{} simulations are run for GR, F5 and N1.2, with identical initial conditions as described in Section~\ref{sec:validation_sim_specs}. The \FML{} force resolution is chosen to be $N_{\mathrm{g}}=2N_{\mathrm{p}}^{\mathrm{1D}}$, given that our comparison to the {\it N}-body results from \taka{} in Section~\ref{sec:validation} shows that increasing $N_{\mathrm{g}}$ beyond this does not significantly improve the accuracy of the statistical features of the predicted convergence field. We again use 40 timesteps between $z_{\rm ini}=20$ to $z=0$, as in Section~\ref{sec:validation_sim_specs}, and all lightcone specifications ({\it i.e.}, $z_*$, $\Delta a$, number of box replicas etc.) remain the same as before. We compare the statistics in MG to the errors on the GR predictions obtained with $10$ {\it N}-body full-sky maps \citep{takahashi_full-sky_2017}, and hence show any detectable signatures of MG beyond these uncertainties. 

To model structure formation under $f(R)$ gravity, \FML{} has an additional parameter called the screening efficiency (see Eq. 2.6 in \citealt{fiorini_fast_2021}), which determines how effectively the enhancement of gravity is screened in high-density regions. This parameter needs to be calibrated, and we first do this by comparing the power spectrum predicted by \FML{} to the theoretical prediction for F5, as will be described in Section~\ref{sec:extension_to_MG_power_spectrum}. After tuning the screening efficiency, we find that a value of $f_s=3.0$ gives the most accurate $C_{\ell}^{\kappa\kappa}$ up to the scale at which we can trust \FML{} ($\ell \sim 750$, as discussed in Section~\ref{sec:validation_power_spectrum}). 

Given the agreement between \FML{} and \taka{} presented in Section~\ref{sec:validation}, we downgrade the pixel resolution of the convergence map to $N_{\mathrm{side}}=256$ when considering the PDF and convergence peak counts, and to $N_{\mathrm{side}}=128$ for the MFs. The corresponding smoothing scale by which these maps are smoothed are $\theta_G=19.93$ arcmins and $\theta_G=39.85$ arcmins, respectively (see Table \ref{tab:smoothing_scale_MG}). As before, we normalise the convergence maps by the pixel standard deviation, $\sigma_{\rm pix}$, which is provided for GR, N1.2 and F5 for a given $N_{\mathrm{side}}$ in Table \ref{tab:smoothing_scale_MG}. We see a larger pixel standard deviation for both F5 and N1.2 as compared to GR, which is expected given the enhancement of gravity in these models. 

\begin{table}  
\caption{Smoothing scale in arcmins, $\theta_G$, and pixel standard deviation, $\sigma_{\rm pix}$, of the convergence maps predicted with \FML{} under GR, F5 and N1.2 gravity, for the corresponding pixel resolution, $N_{\rm side}$. When considering the PDF, peak counts and MFs, we downgrade $N_{\mathrm{side}}$ to ensure sufficient accuracy, following the validation tests presented in Section~\ref{sec:validation}. The maps are then smoothed by $\theta_G$ and normalised by $\sigma_{\rm pix}$, as described in Section~\ref{sec:validation_sim_specs}. The value of $\theta_G$ reported here is the same as in Table \ref{tab:smoothing_scale}:  we repeat it for clarity for the two values of $N_{\mathrm{side}}$ that are used.  }
\centering
 \begin{tabular}{||c|c|c|c|c||} 
 \hline
 $N_{\rm side}$ & $\theta_G$ (arcmins) & $\sigma_{\rm pix}^{\rm GR}$ & $\sigma_{\rm pix}^{\rm F5}$ & $\sigma_{\rm pix}^{\rm N1.2}$ \\ [0.5ex] 
 \hline\hline
 128 & 39.85 & 0.00418 & 0.00425 & 0.00438\\ 
 256 & 19.93 & 0.00559 & 0.00573 & 0.00588 \\ [1ex] 
 \hline
 \end{tabular}
 \label{tab:smoothing_scale_MG}
\end{table}

\subsection{Power spectrum}\label{sec:extension_to_MG_power_spectrum}

\begin{figure}
    \centering
    \includegraphics[width=\linewidth]{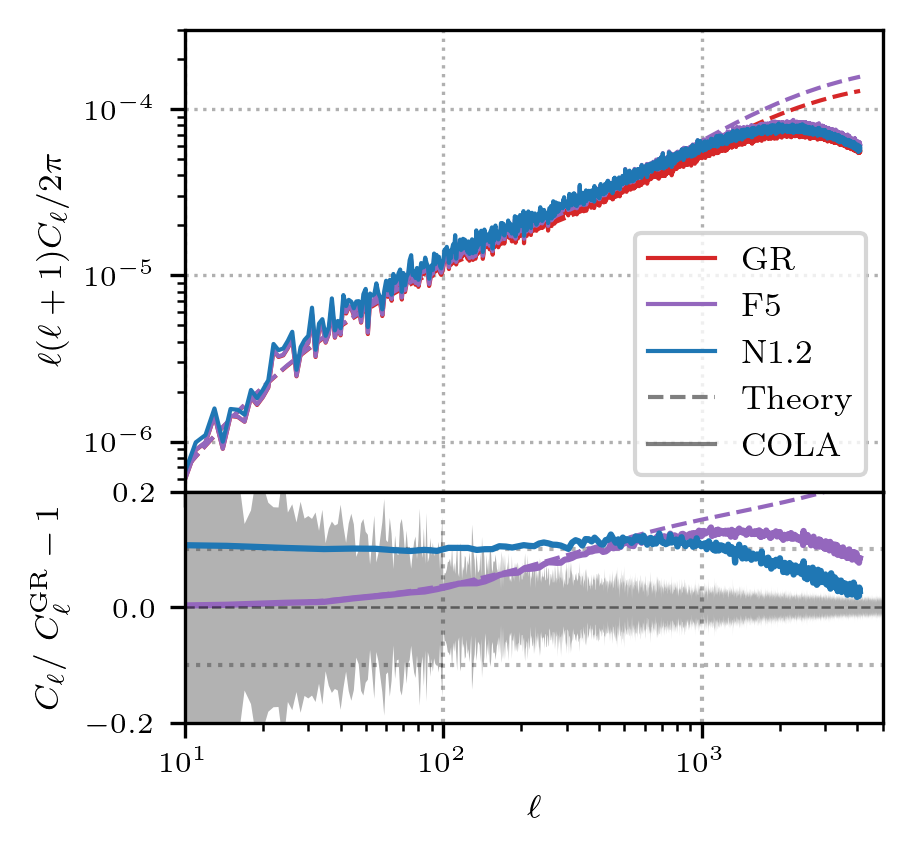}
    \caption{Angular power spectrum of the $z=0$ convergence field computed with \FML{} under GR, F5 and N1.2 gravity (solid black, blue and red lines, respectively). Dashed lines show theoretical predictions computed by integrating the non-linear matter power spectrum, $P(k,z)$, along the line of sight, as described in Section~\ref{sec:extension_to_MG_power_spectrum} (in GR, the \textsc{Halofit} $P(k,z)$ is obtained with \textsc{CAMB} \citep{lewis_efficient_2000}, and in F5, we multiply the \textsc{Halofit} $P(k,z)$ with the MG boost obtained with \textsc{eMantis} \citep{saez-casares_e-mantis_2023}). The choice of screening efficiency in F5 is chosen to be $f_s=3.0$ based on the comparison with theory up to the scale $\ell\sim750$ at which we can trust the \FML{} predictions (Section~\ref{sec:validation_power_spectrum}). There is a scale-dependent enhancement of power in F5, while N1.2 shows a constant enhancement with respect to GR across all scales. Screening ensures that the GR predictions are recovered for higher $\ell$. The shaded region corresponds to the errors on the GR prediction, estimated by the standard deviation of $10$ \taka{} realisations.}
    \label{fig:Cls_MG_v_GR}
\end{figure}

Figure~\ref{fig:Cls_MG_v_GR} shows the weak lensing angular power spectrum, $C_{\ell}^{\kappa\kappa}$, computed on-the-fly with \FML{} under GR, F5 and N1.2 gravity, following the method described in Section~\ref{sec:WL_map_construction}. The GR and F5 results are also compared to their corresponding theoretical predictions, which are computed by integrating the non-linear matter power spectrum, $P(k,z)$, from $z_*=1.0334$ to $z=0$ according to Eq.~\ref{eq:C(ell)}. To compute the GR prediction, we start from the GR non-linear matter power spectrum $P^{\rm GR}$,  as described in Section~\ref{sec:validation_power_spectrum}. To compute the F5 prediction, we obtain the MG boost, $P^{\rm F5} / P^{\rm GR}$, using the \textsc{eMantis}\footnote{\url{https://zenodo.org/records/13900122}} emulator \citep{saez-casares_e-mantis_2023}, for the chosen cosmology. Multiplying the \textsc{Halofit} and \textsc{eMantis} output $P^{\rm GR} \times P^{\rm F5} / P^{\rm GR}$ then yields the nonlinear matter power spectrum in F5. We then integrate this according to Eq.~\ref{eq:C(ell)} to obtain the theoretical prediction of the weak lensing angular power spectrum in $F5$. 

We run several \FML{} simulations with varying screening efficiencies and obtain the angular power spectrum that \FML{} computes on the fly. By comparing the \FML{}-predicted $C_{\ell}^{\kappa\kappa}$ in F5 to its theoretical prediction, we first tune the screening efficiency, $f_s$, of \FML{}. We find that a value of $f_s=3.0$ gives the best agreement up to $\ell\sim750$, as we can see in Figure \ref{fig:Cls_MG_v_GR}, which is the scale at which the accuracy of \FML{} breaks down compared to the full {\it N}-body results (Section~\ref{sec:validation_power_spectrum}). This screening efficiency is used throughout the remainder of this work to compare the convergence field statistics in F5 to GR.

The convergence power spectrum contains information about the scale-dependent linear growth rate in $f(R)$ gravity, and the environmental dependence of non-linear gravitational growth due to screening. The power spectrum is enhanced by a factor of $\sim5\%$ at $\ell \sim100$ and $\sim10\%$ at $\ell \sim1000$, and the screening mechanism acts to recover GR at $\ell\gtrsim1000$. N1.2 exhibits the expected scale-independent enhancement of power with respect to GR, of approximately $10\%$ between $10< \ell <1000$. Figure \ref{fig:Cls_MG_v_GR} shows that both F5 and N1.2 would be distinguishable from GR using the convergence field angular power spectrum at $\ell \gtrsim 400$, consistent with previous forecasts that demonstrate the power spectrum's ability to constrain $f_{R0}$ and $r_c$ \citep{schneider_baryonic_2020, harnois-deraps_mglens_2022, mancini_degeneracies_2023, collaboration_euclid_2024_2, tsedrik_stage-iv_2024}.

\subsection{Bispectrum}\label{sec:extension_to_MG_bispectrum}
\begin{figure*} 
  \centering
  \includegraphics[width=\textwidth]{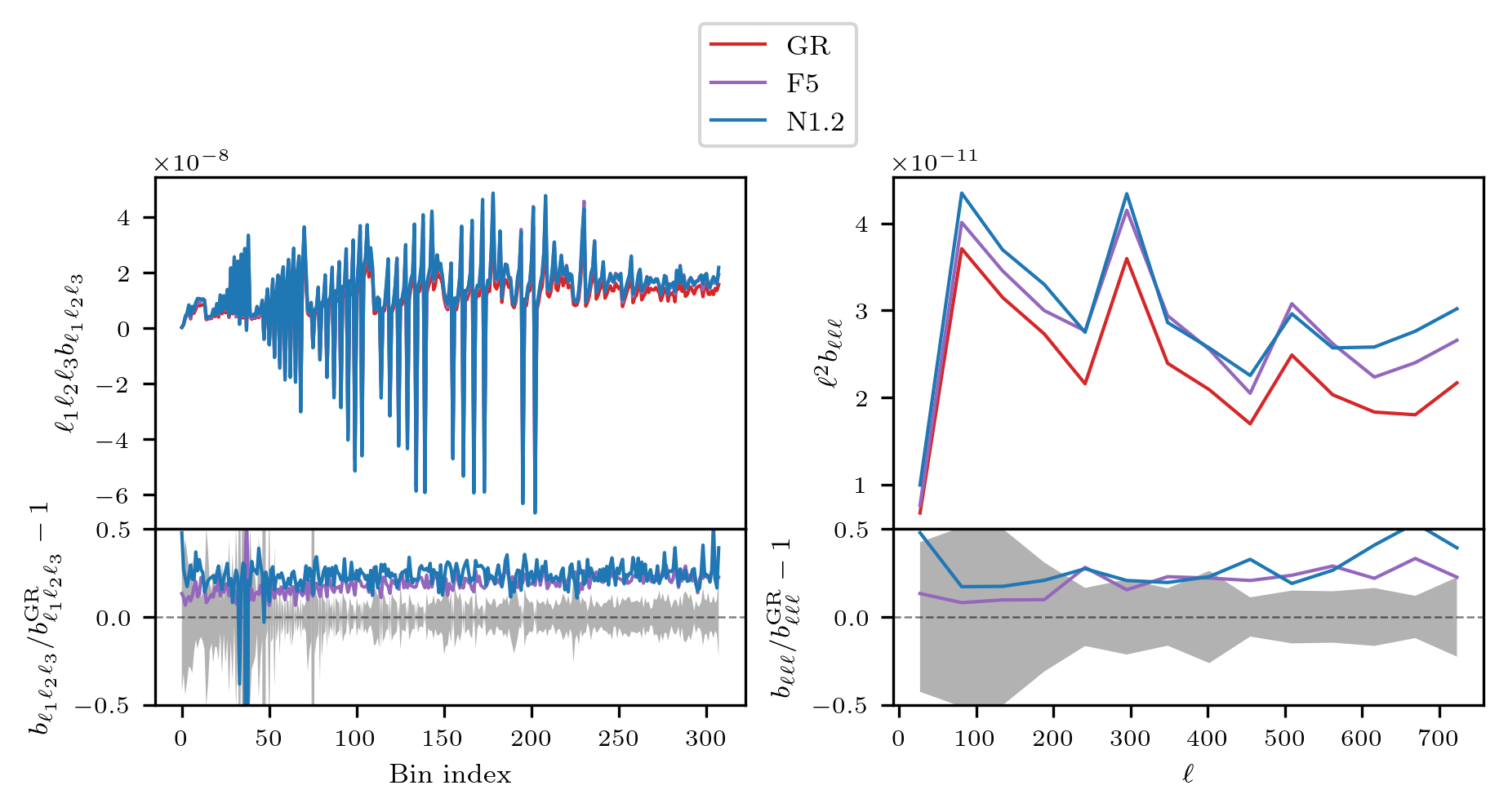}
  \caption{Bispectrum of the $z=0$ convergence field computed with \FML{} under GR, F5 and N1.2 gravity (black, blue and red lines, respectively). The shaded region corresponds to the errors on the GR prediction, estimated by the standard deviation of $10$ \taka{} realisations. The bispectrum is computed using a total number of bins $N_{\mathrm{bins}}=15$ within $10<\ell<750$. In the left panel, we show all $\ell_1\ell_2\ell_3$ configurations satisfying the triangle condition, whereas in the right panel, we show the equilateral configurations ($\ell \equiv \ell_1=\ell_2=\ell_3$). We see an enhancement in the bispectrum for both F5 and N1.2, as compared to GR.}
  \label{fig:bispectrum_MG_v_GR}
\end{figure*}

The bispectrum is again computed using \textsc{PolySpec} \citep{philcox_optimal_2023}, following the method described in Section~\ref{sec:validation_bispectrum}. We compute the bisepctrum for $N_{\mathrm{bins}}=15$ bins between $10 < \ell < 750$. The maximum $\ell$ value up to which we compute the bispectrum was chosen according to our validation tests of the \FML{} convergence field using the power spectrum (Section~\ref{sec:validation_power_spectrum}). 

Figure \ref{fig:bispectrum_MG_v_GR} shows a comparison of the bispectra predicted under GR, F5 and N1.2 gravity using \FML{}. Increasing bin index corresponds to increasing $\ell_3$ first, then $\ell_2$, then $\ell_1$, taking all $\ell_1,\ell_2,\ell_3$ configurations that satisfy the triangle condition ($\ell_1 \leq \ell_2 + \ell_3$ for $\ell_1 \geq \ell_2, \ell_3$), as described in Section~\ref{sec:validation_power_spectrum}.

Results are shown for all configurations (left panel) and equilateral configurations (right panel) (\textit{i.e.,} $\ell \equiv \ell_1=\ell_2=\ell_3$), where we instead plot with respect to $\ell$. We find an increase in the bispectrum of approximately $20\%$ for both F5 and N1.2, irrespective of the triangle configuration considered. This change is predicted to be beyond the statistical uncertainty of the convergence field bispectrum, estimated from $10$ realisations of full sky maps in GR, suggesting that both F5 and N1.2 would be detectable using the bispectrum up to $\ell\sim750$. This is consistent with the findings of \cite{shirasaki_imprint_2017}. To explore potential degeneracies with cosmological parameters, we ran an additional GR simulation with an increased value of $\sigma_8$. We found that the bispectrum amplitudes increase in a similar way to those in the MG cases, suggesting that $\sigma_8$ is degenerate with MG when considering the bispectrum alone. However, a comparison of the convergence power spectra suggests that a joint analysis of the power spectrum and bispectrum may be able to break this degeneracy \citep{shirasaki_imprint_2017}.

\subsection{PDF}\label{sec:extension_to_MG_PDF}
\begin{figure}
    \centering
    \includegraphics[width=\linewidth]{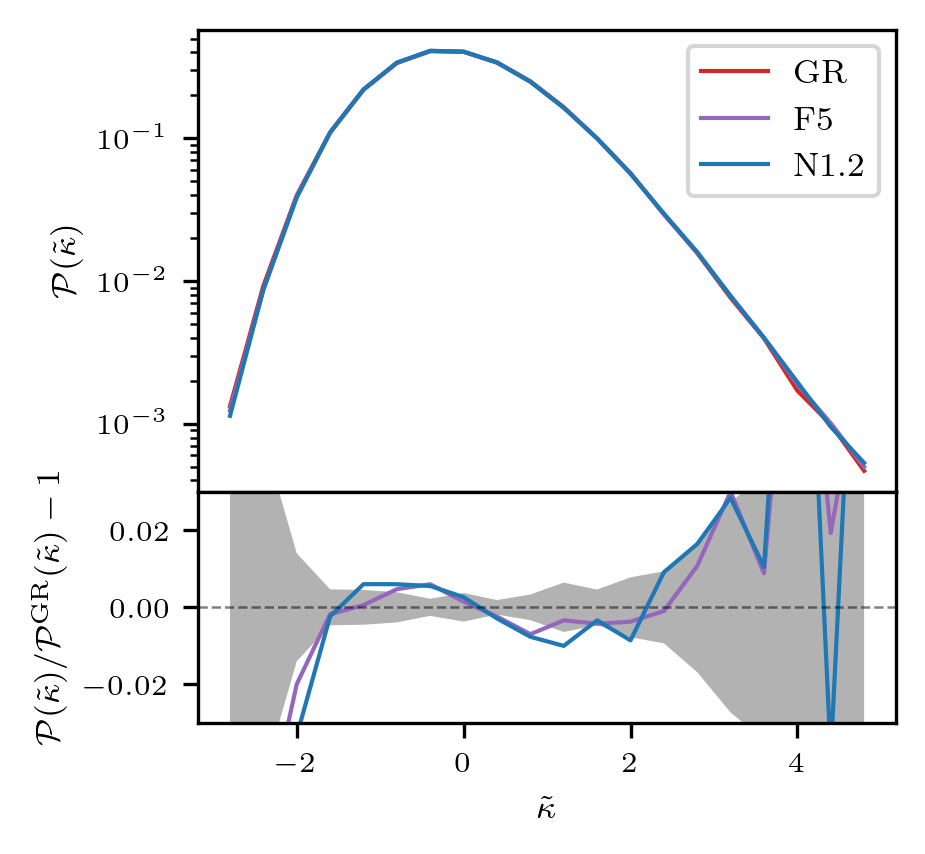}
    \caption{Pixel PDF, $\mathcal{P}$, of the $z=0$ convergence field, $\Tilde{\kappa}$, computed with \FML{} under GR, F5 and N1.2 gravity (black, blue and red lines, respectively). All convergence maps have been downgraded to a pixel resolution of $N_{\rm side}=256$, following the validation test presented in Section~\ref{sec:validation_PDF}. The shaded area corresponds to the errors on the GR prediction, estimated by the standard deviation of $10$ \taka{} realisations. Both the F5 and N1.2 predictions show a deviation from GR beyond the estimated statistical uncertainties around $\Tilde{\kappa} \sim \pm1$. }
    \label{fig:PDF_MG_v_COLA}
\end{figure}

Figure~\ref{fig:PDF_MG_v_COLA} shows a comparison of the $z=0$ convergence field PDF predicted with \FML{} under GR, F5 and N1.2 gravity. Following the validation tests in Section~\ref{sec:validation_PDF}, we set the pixel resolution to $N_{\mathrm{side}}=256$, and smooth the $\Tilde{\kappa}$-maps by a corresponding scale of 19.93 arcmins (see Table \ref{tab:smoothing_scale_MG}). The PDF is again computed using a histogram of $\Tilde{\kappa}$ in 21 bins between $-3 < \Tilde{\kappa} < 5$, according to Eq.~\ref{eq:PDF_hist}, where the total number of pixels is $N_{\mathrm{\rm pix}}=12N_{\mathrm{side}}^2=786,432$ and the binning width is $\Delta \Tilde{\kappa} \sim0.38$. 

The PDFs of the F5 and N1.2 convergence maps show a decrease for negative $\Tilde{\kappa}$ and a increase for positive $\Tilde{\kappa}$ as compared with GR, as expected from the enhanced gravity in these models. Based on the estimated statistical uncertainties from 10 \taka{} realisations, we find that deviations from GR are detectable for $-1 < \Tilde{\kappa} < 1$. This confirms that the one-point PDF is sensitive to MG \citep{cataneo_matter_2022, gough_one-point_2022}. Although significantly lower pixel resolution for \FML{} is required compared to the power spectrum and bispectrum, it still provides a useful way to distinguish modified gravity models from GR.

\subsection{Peak counts}\label{sec:extension_to_MG_peaks}

Next, we examine the imprints of MG in the peak statistics of the convergence field. Following Section~\ref{sec:validation_peaks}, we downgrade the $\Tilde{\kappa}$-maps to a pixel resolution of $N_{\mathrm{side}}=256$, and the maps are therefore smoothed with Gaussian kernel of width 19.93 arcmins (see Table \ref{tab:smoothing_scale_MG}). Peak counts are again computed using \textsc{LensTools} \citep{petri_mocking_2016}, which follows the method described in Section~\ref{sec:validation_peaks}. We use a total of 21 bins between $-3 < \Tilde{\kappa} < 5$. 

\begin{figure}
    \centering
    \includegraphics[width=\linewidth]{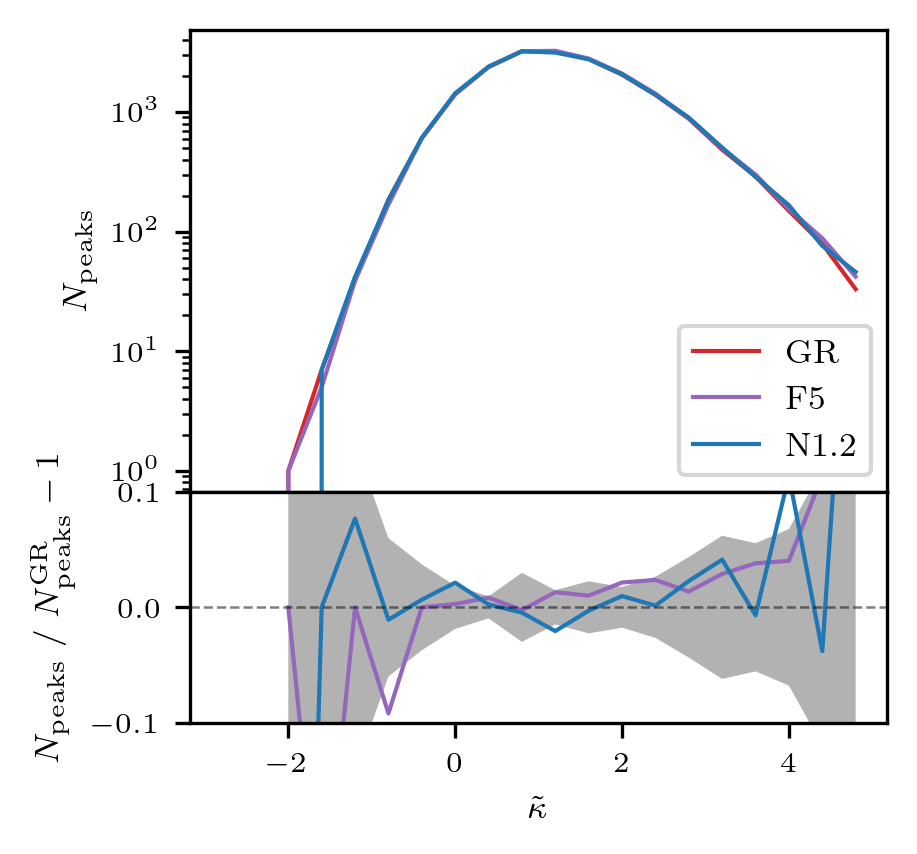}
    \caption{Peak counts, $N_{\rm peaks}$, of the $z=0$ convergence field, $\Tilde{\kappa}$, computed with \FML{} under GR, F5, N1.2 gravity (black, blue and red lines, respectively). $\Tilde{\kappa}$-maps are downgraded to $N_{\rm side}=256$ following the \FML{} validation tests in Section~\ref{sec:validation_peaks}. The shaded region corresponds to the errors on the GR prediction, estimated by the standard deviation of $10$ \taka{} realisations. Although we see an enhancement in $N_{\rm peaks}$ for $\Tilde{\kappa}>2$ in both F5 and N1.2, these effects would be indistinguishable beyond our estimated uncertainties on the GR prediction. }
    \label{fig:peakcount}
\end{figure}

Figure \ref{fig:peakcount} shows that the difference between GR and MG is larger for larger $\Tilde{\kappa}$. Due to the low resolution of the map ($N_{\mathrm{side}}=256$), these deviations are only marginally detectable compared to the statistical uncertainties estimated from $10$ \taka{} realisations in GR. The convergence peak count was previously used to place constraints of $f_{R0} < 10^{-5.16}$ from the Canada-France-Hawaii Telescope Lensing Survey \citep{liu_constraining_2016}, so although F5 is not distinguishable from GR given the estimated errors in this work, the slight enhancement of $N_{\mathrm{peaks}}$ is qualitatively consistent with previous works \citep{higuchi_imprint_2016, shirasaki_imprint_2017}. 

\subsection{Minkowski Functionals}\label{sec:extension_to_MG_MFs}
\begin{figure*} 
  \centering
  \includegraphics[width=\linewidth]{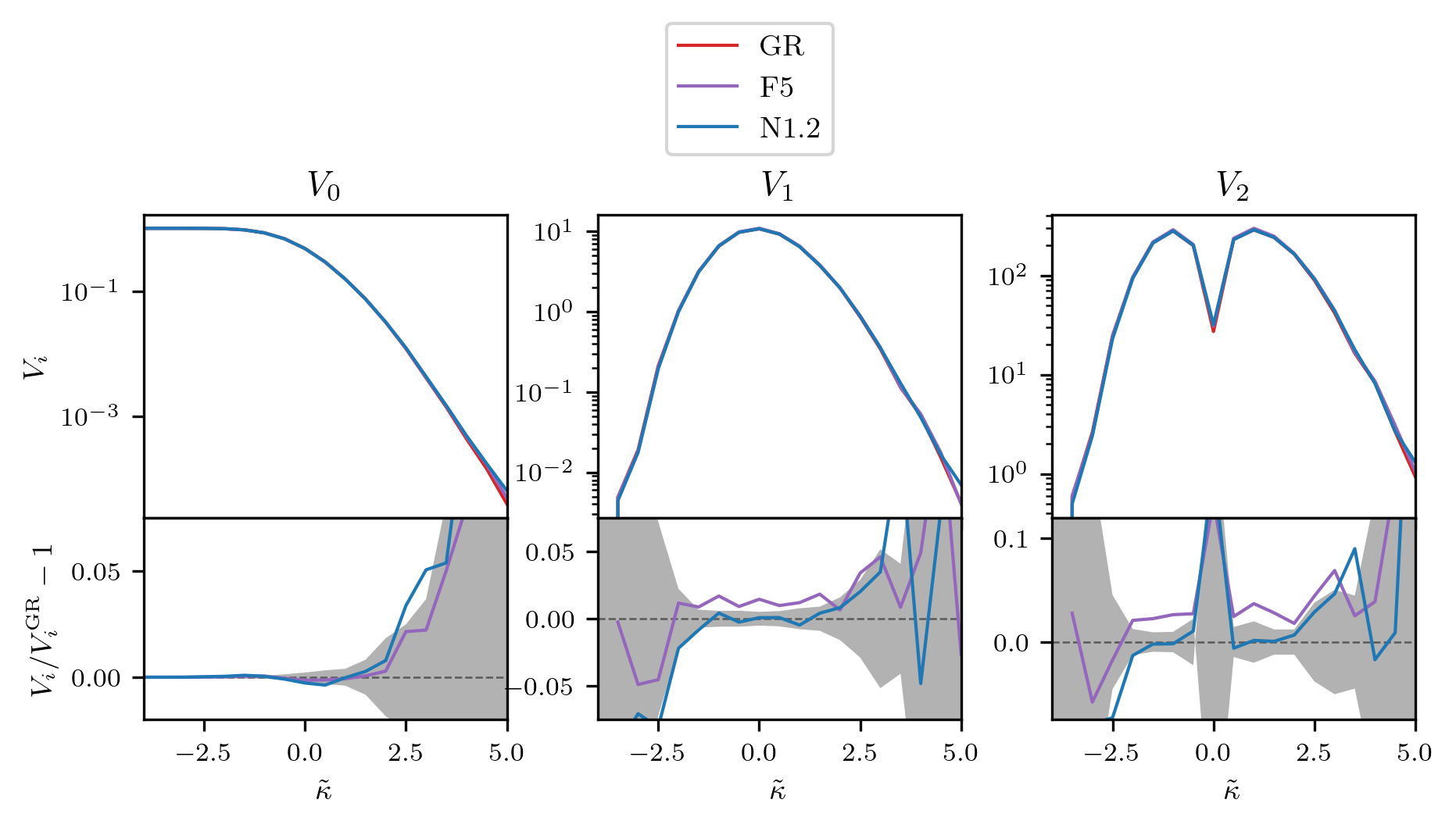}
  \caption{Minkowksi functionals of the $z=0$ convergence fields, $\Tilde{\kappa}$, computed with \FML{} under GR, F5 and N1.2 gravity (black, blue and red lines, respectively): $V_0$ (left panel); $V_1$ (middle panel); $V_2$ (right panel). All maps have been downgraded to a pixel resolution $N_{\rm side}=128$, as this is the resolution that most accurately captures the morphological features contained within the $\Tilde{\kappa}$-maps in GR (Section~\ref{sec:validation_MFs}). The corresponding smoothing scale that is applied is 39.85 arcmin (Table \ref{tab:smoothing_scale_MG}). The shaded region corresponds to the errors on the GR prediction, estimated by the standard deviation of $10$ \taka{} realisations. For N1.2 gravity, we see no deviation from GR beyond the estimated statistical uncertainties for all MFs, while F5 gravity would be distinguishable from GR in $V_1$ within the range $-1 \lesssim \Tilde{\kappa} \lesssim 1$, and in $V_2$ within the range $-2 \lesssim \Tilde{\kappa} \lesssim 3$. }
  \label{fig:MFs_MG_v_GR}
\end{figure*}
  
We compute the MFs of the $\Tilde{\kappa}$-maps after setting the pixel resolution to $N_{\mathrm{side}}=128$, given the results in Section~\ref{sec:validation_MFs}. The corresponding smoothing scale that is applied is 39.85 arcmin, as shown in Table \ref{tab:smoothing_scale_MG}. We again use \textsc{Pynkowski} \citep{carones_minkowski_2023} to compute the MFs in 21 bins between $-5 < \Tilde{\kappa} < 5$, following the method described in Section~\ref{sec:validation_MFs}.

Figure \ref{fig:MFs_MG_v_GR} shows a comparison of the convergence field MFs predicted in GR, F5 and N1.2. When considering $V_0$, which describes the area of regions with a convergence higher than a certain threshold, both F5 and N1.2 predictions show a lower value than GR for $0 \lesssim \Tilde{\kappa} \lesssim 1$, but a higher value than GR for $\Tilde{\kappa} \gtrsim 2$. This is due to the enhanced strength of gravity, whereby low density regions in GR become lower density, and hence have a lower convergence, while enhanced structure in high density regions leads to higher convergence. 
$V_1$ and $V_2$ encode more information about the topology of the convergence field, where we do not see significant deviations between N1.2 and GR predictions. However, we see that F5 would be detectable beyond the statistical uncertainties when considering both $V_1$ and $V_2$, where there is difference of $\sim2.5\%$ for $-1 \lesssim \Tilde{\kappa} \lesssim 1$ and $-2 \lesssim \Tilde{\kappa} \lesssim 3$, respectively. Our results are consistent with previous studies of MFs under $f(R)$ gravity \citep{ling_distinguishing_2015, shirasaki_imprint_2017, jiang_minkowski_2024}.

\subsection{Computational resources}\label{sec:extension_to_MG_resources}
To conclude this section, we present the computational resources required to predict the $z=0$ weak lensing convergence field with \FML{} under GR, F5 and N1.2 gravity. The simulation specifications ({\it i.e.,} mass resolution and timesteps) and lightcone specifications ({\it i.e.,} source redshift and radial binning width) are described in Section~\ref{sec:validation_sim_specs}. Following our validation tests of the lightcone implementation in \FML{}, we choose the force resolution of each \FML{} run to be $N_{\mathrm{g}}=2N_{\mathrm{p}}^{\mathrm{1D}}$, as described in Section~\ref{sec:extension_to_MG_sim_specs}.

All simulations were run on a total of 128 Intel Xeon 2.1Ghz CPUs. Table \ref{tab:comp_resources} shows the total CPU hours for each of the three \FML{} simulations.

\begin{table}  
\caption{Computational resources used to obtain the $z=0$ convergence field with \FML{} under GR, F5 and N1.2 gravity.}
\centering
 \begin{tabular}{||c|c||} 
 \hline
 Model & CPU hours  \\ [0.5ex] 
 \hline\hline
 GR & 317.53   \\ 
 $f(R)$ & 404.96 \\ 
 nDGP & 514.84  \\      [1ex] 
 \hline
 \end{tabular}
 \label{tab:comp_resources}
\end{table}

\section{Discussion \& Conclusions}\label{sec:discussion}

In this work, we have extended the \FML{} library to generate weak lensing convergence maps via lightcone construction. This method allows us to predict the convergence field in both GR and MG at a reduced computational cost than full {\it N}-body simulations. To assess the accuracy of the convergence maps generated with our lightcone implementation into \FML{}, we carried out a detailed comparison against the full sky maps of \taka{}, which were generated with full {\it N}-body simulations in GR. For this validation, we have considered five summary statistics: the power spectrum, bispectrum, probability distribution function, peak counts and Minkowski functionals, all of which are sensitive to non-linear structure formation. 

For each statistic, we investigate whether increasing the \FML{} force resolution, $N_{\mathrm{g}}$, enhances the accuracy of the predicted convergence maps. We find that increasing $N_{\mathrm{g}}$ beyond $N_{\mathrm{g}}=2N_{\mathrm{p}}^{\mathrm{1D}}$ does not yield a higher accuracy when using a fixed number of 40 timesteps. 

We found that the \FML{} convergence power spectrum is accurate to within $5\%$ ($10\%$) of the {\it N}-body results up to $\ell \sim 500$ ($\ell \sim 750$). We can therefore trust the \FML{}-generated convergence maps up to a scale of $\ell_{\rm max}\sim750$. The agreement is worse for the bispectrum, where we lose power at small scales due to the limitation of approximate methods, giving an agreement of $25\%$ at $\ell\sim700$ for equilateral configurations. For the PDF, peak counts and MFs, we examined the pixel resolution, $N_{\mathrm{side}}$, at which we can trust the \FML{}-generated maps. We find that $N_{\mathrm{side}}=256$ is required to accurately reproduce the PDF and peak counts within the errors estimated from $10$ \taka{} realisations, while a slightly coarser resolution of $N_{\mathrm{side}}=128$ is required to capture the morphological properties of the convergence field within the MFs. 

These validation tests provide a baseline for extending our analyses to MG models, which is a main purpose of implementing the lightcone generation within \FML{}. With \FML{}, we can easily model non-linear structure formation under many theories of MG, and the new lightcone implementation allows for a self-consistent comparison of statistical features within the GR and MG convergence fields. All results therefore focus on the ratio between the convergence statistics in MG (namely F5 and N1.2) and GR. The statistics are computed on the convergence maps with the optimal specifications found with our \FML{} validation tests. 

We estimated the uncertainty of each statistic using $10$ realisations of the full sky maps of \taka{}. These uncertainties are only indicative. However, they allow us to compare the ability of each statistic to distinguish between GR and MG. We see an enhancement in the convergence power spectrum of $\sim 10\%$ at $\ell \sim 750$ for both F5 and N1.2, and an enhancement in the bispectrum of $\sim 20\%$, irrespective of which $(\ell_1,\ell_2,\ell_3)$ configuration is used, which agrees with previous work. 
We found that both F5 and N1.2 gravity show distinct deviations from GR for the PDF, peak counts and MFs, as expected from enhanced gravity. In particular, the peak counts and MFs provide promising ways to distinguish between F5 and GR, as was found in {\it e.g.} \cite{shirasaki_imprint_2017}. Even though we need to use lower pixel resolution to predict beyond $n$-point statistics using \FML{}, they still have an ability to discriminate MG from GR. A more thorough analysis including accurate covariance estimate (requiring $\gtrsim$100 realisations) and realistic noise modelling is left to future work.

The implementation presented in this work provides a computationally efficient tool to generate a large number of convergence maps under different theories of gravity. By enabling accurate predictions of the convergence field, and allowing for the fast creation of mock data sets, this work is a step towards making more detailed, and stringent, comparisons between our theoretical models and cosmological surveys, beyond the power spectrum.
Our results additionally highlight the potential of weak lensing data to detect signatures of MG, particularly through higher-order statistics, as was shown in previous works. Future efforts will focus on mock generation to place constraints on deviations from GR with weak lensing data, and exploring a combination of weak lensing and galaxy clustering for a more comprehensive view of our Universe. 

\section*{Acknowledgements}
SH is supported by the UK Science and Technology Facilities Council (STFC) grant number ST/X508688/1 and funding from the University of Portsmouth. SH was also supported by an STFC Long-term Attachment grant number ST/W507738/1, to visit the University of Oslo. DS and KK are supported by STFC  grant number ST/W001225/1. AI was supported by STFC grant number ST/S000550/1.

SH thanks David Mota for helpful discussions while visiting the University of Oslo.

Numerical computations were carried out on the Sciama High Performance Computing (HPC) cluster, which is supported by the Institute of Cosmology and Gravitation, the South-East Physics Network (SEPNet) and the University of Portsmouth.

SH dedicates this paper to the memory of her sister, Liana Hoyland, and the significance of the day of submission.

For the purpose of open access, we have applied a Creative Commons Attribution (CC BY) licence to any Author Accepted Manuscript version arising.

\section*{Data Availability}

The supporting research data and code are available upon reasonable request from the corresponding author.


\bibliographystyle{mnras}
\bibliography{references}

\begin{thebibliography}{}
\makeatletter
\relax
\def\mn@urlcharsother{\let\do\@makeother \do\$\do\&\do\#\do\^\do\_\do\%\do\~}
\def\mn@doi{\begingroup\mn@urlcharsother \@ifnextchar [ {\mn@doi@} {\mn@doi@[]}}
\def\mn@doi@[#1]#2{\def\@tempa{#1}\ifx\@tempa\@empty \href {http://dx.doi.org/#2} {doi:#2}\else \href {http://dx.doi.org/#2} {#1}\fi \endgroup}
\def\mn@eprint#1#2{\mn@eprint@#1:#2::\@nil}
\def\mn@eprint@arXiv#1{\href {http://arxiv.org/abs/#1} {{\tt arXiv:#1}}}
\def\mn@eprint@dblp#1{\href {http://dblp.uni-trier.de/rec/bibtex/#1.xml} {dblp:#1}}
\def\mn@eprint@#1:#2:#3:#4\@nil{\def\@tempa {#1}\def\@tempb {#2}\def\@tempc {#3}\ifx \@tempc \@empty \let \@tempc \@tempb \let \@tempb \@tempa \fi \ifx \@tempb \@empty \def\@tempb {arXiv}\fi \@ifundefined {mn@eprint@\@tempb}{\@tempb:\@tempc}{\expandafter \expandafter \csname mn@eprint@\@tempb\endcsname \expandafter{\@tempc}}}

\bibitem[\protect\citeauthoryear{Abbott et~al.,}{Abbott et~al.}{2017}]{abbott_gw170817_2017}
Abbott B.,  et~al., 2017, \mn@doi [Physical Review Letters] {10.1103/PhysRevLett.119.161101}, 119, 161101

\bibitem[\protect\citeauthoryear{Abdalla et~al.,}{Abdalla et~al.}{2022}]{abdalla_cosmology_2022}
Abdalla E.,  et~al., 2022, \mn@doi [Journal of High Energy Astrophysics] {10.1016/j.jheap.2022.04.002}, 34, 49

\bibitem[\protect\citeauthoryear{Albrecht et~al.,}{Albrecht et~al.}{2006}]{albrecht_report_2006}
Albrecht A.,  et~al., 2006, Report of the {Dark} {Energy} {Task} {Force}, \mn@doi{10.48550/arXiv.astro-ph/0609591}, \url {http://arxiv.org/abs/astro-ph/0609591}

\bibitem[\protect\citeauthoryear{Arnold, Fosalba, Springel, Puchwein  \& Blot}{Arnold et~al.}{2019}]{arnold_modified_2019}
Arnold C.,  Fosalba P.,  Springel V.,  Puchwein E.,   Blot L.,  2019, \mn@doi [Monthly Notices of the Royal Astronomical Society] {10.1093/mnras/sty3044}, 483, 790

\bibitem[\protect\citeauthoryear{Baker et~al.,}{Baker et~al.}{2021}]{baker_novel_2021}
Baker T.,  et~al., 2021, \mn@doi [Reviews of Modern Physics] {10.1103/RevModPhys.93.015003}, 93, 015003

\bibitem[\protect\citeauthoryear{Baldauf, Seljak  \& Senatore}{Baldauf et~al.}{2011}]{baldauf_primordial_2011}
Baldauf T.,  Seljak U.,   Senatore L.,  2011, \mn@doi [Journal of Cosmology and Astroparticle Physics] {10.1088/1475-7516/2011/04/006}, 2011, 006

\bibitem[\protect\citeauthoryear{Bartelmann \& Schneider}{Bartelmann \& Schneider}{2001}]{bartelmann_weak_2001}
Bartelmann M.,  Schneider P.,  2001, \mn@doi [Physics Reports] {10.1016/S0370-1573(00)00082-X}, 340, 291

\bibitem[\protect\citeauthoryear{Barthelemy, Codis, Uhlemann, Bernardeau  \& Gavazzi}{Barthelemy et~al.}{2020}]{barthelemy_nulling_2020}
Barthelemy A.,  Codis S.,  Uhlemann C.,  Bernardeau F.,   Gavazzi R.,  2020, \mn@doi [Monthly Notices of the Royal Astronomical Society] {10.1093/mnras/staa053}, 492, 3420

\bibitem[\protect\citeauthoryear{Bernardeau, Colombi, Gaztanaga  \& Scoccimarro}{Bernardeau et~al.}{2002}]{bernardeau_large-scale_2002}
Bernardeau F.,  Colombi S.,  Gaztanaga E.,   Scoccimarro R.,  2002, \mn@doi [Physics Reports] {10.1016/S0370-1573(02)00135-7}, 367, 1

\bibitem[\protect\citeauthoryear{Boruah \& Rozo}{Boruah \& Rozo}{2023}]{boruah_map-based_2023}
Boruah S.~S.,  Rozo E.,  2023, Map-based cosmology inference with weak lensing -- information content and its dependence on the parameter space, \mn@doi{10.48550/arXiv.2307.00070}, \url {http://arxiv.org/abs/2307.00070}

\bibitem[\protect\citeauthoryear{Boyle, Uhlemann, Friedrich, Barthelemy, Codis, Bernardeau, Giocoli  \& Baldi}{Boyle et~al.}{2021}]{boyle_nuw_2021}
Boyle A.,  Uhlemann C.,  Friedrich O.,  Barthelemy A.,  Codis S.,  Bernardeau F.,  Giocoli C.,   Baldi M.,  2021, \mn@doi [Monthly Notices of the Royal Astronomical Society] {10.1093/mnras/stab1381}, 505, 2886

\bibitem[\protect\citeauthoryear{Brando, Fiorini, Koyama  \& Winther}{Brando et~al.}{2022}]{brando_enabling_2022}
Brando G.,  Fiorini B.,  Koyama K.,   Winther H.~A.,  2022, \mn@doi [Journal of Cosmology and Astroparticle Physics] {10.1088/1475-7516/2022/09/051}, 2022, 051

\bibitem[\protect\citeauthoryear{Carones, CarrónDuque, Marinucci, Migliaccio  \& Vittorio}{Carones et~al.}{2023}]{carones_minkowski_2023}
Carones A.,  CarrónDuque J.,  Marinucci D.,  Migliaccio M.,   Vittorio N.,  2023, \mn@doi [Monthly Notices of the Royal Astronomical Society] {10.1093/mnras/stad3002}, 527, 756

\bibitem[\protect\citeauthoryear{Cataneo, Uhlemann, Arnold, Gough, Li  \& Heymans}{Cataneo et~al.}{2022}]{cataneo_matter_2022}
Cataneo M.,  Uhlemann C.,  Arnold C.,  Gough A.,  Li B.,   Heymans C.,  2022, \mn@doi [Monthly Notices of the Royal Astronomical Society] {10.1093/mnras/stac904}, 513, 1623

\bibitem[\protect\citeauthoryear{Clifton, Ferreira, Padilla  \& Skordis}{Clifton et~al.}{2012}]{clifton_modified_2012}
Clifton T.,  Ferreira P.~G.,  Padilla A.,   Skordis C.,  2012, \mn@doi [Physics Reports] {10.1016/j.physrep.2012.01.001}, 513, 1

\bibitem[\protect\citeauthoryear{Cooray \& Hu}{Cooray \& Hu}{2000}]{cooray_weak_2000}
Cooray A.,  Hu W.,  2000, Weak {Gravitational} {Lensing} {Bispectrum}, \url {http://arxiv.org/abs/astro-ph/0004151}

\bibitem[\protect\citeauthoryear{Creminelli \& Vernizzi}{Creminelli \& Vernizzi}{2017}]{creminelli_dark_2017}
Creminelli P.,  Vernizzi F.,  2017, \mn@doi [Physical Review Letters] {10.1103/PhysRevLett.119.251302}, 119, 251302

\bibitem[\protect\citeauthoryear{Davies, Harnois-Déraps, Li, Giblin, Hernández-Aguayo  \& Paillas}{Davies et~al.}{2024}]{davies_constraining_2024}
Davies C.~T.,  Harnois-Déraps J.,  Li B.,  Giblin B.,  Hernández-Aguayo C.,   Paillas E.,  2024, Constraining modified gravity with weak lensing peaks, \url {http://arxiv.org/abs/2406.11958}

\bibitem[\protect\citeauthoryear{Ding, Li, Zheng, Luo, Zhang  \& Li}{Ding et~al.}{2024}]{ding_fast_2024}
Ding J.,  Li S.,  Zheng Y.,  Luo X.,  Zhang L.,   Li X.-D.,  2024, \mn@doi [The Astrophysical Journal Supplement Series] {10.3847/1538-4365/ad0c5b}, 270, 25

\bibitem[\protect\citeauthoryear{Dodelson \& Zhang}{Dodelson \& Zhang}{2005}]{dodelson_weak_2005}
Dodelson S.,  Zhang P.,  2005, The {Weak} {Lensing} {Bispectrum}, \url {http://arxiv.org/abs/astro-ph/0501063}

\bibitem[\protect\citeauthoryear{Dvali, Gabadadze  \& Porrati}{Dvali et~al.}{2000}]{dvali_4d_2000}
Dvali G.,  Gabadadze G.,   Porrati M.,  2000, \mn@doi [Physics Letters B] {10.1016/S0370-2693(00)00669-9}, 485, 208

\bibitem[\protect\citeauthoryear{Einasto, Klypin, Hütsi, Liivamägi  \& Einasto}{Einasto et~al.}{2021}]{einasto_evolution_2021}
Einasto J.,  Klypin A.,  Hütsi G.,  Liivamägi L.-J.,   Einasto M.,  2021, \mn@doi [Astronomy \& Astrophysics] {10.1051/0004-6361/202039999}, 652, A94

\bibitem[\protect\citeauthoryear{{Euclid Collaboration: Ajani et al.}}{{Euclid Collaboration: Ajani et al.}}{2023}]{euclid_collaboration_euclid_2023-2}
{Euclid Collaboration: Ajani et al.} 2023, \mn@doi [Astronomy \& Astrophysics] {10.1051/0004-6361/202346017}, 675, A120

\bibitem[\protect\citeauthoryear{{Euclid Collaboration: Koyama et al.}}{{Euclid Collaboration: Koyama et al.}}{2024}]{collaboration_euclid_2024_2}
{Euclid Collaboration: Koyama et al.} 2024, Euclid preparation. {Simulations} and nonlinearities beyond $\Lambda${CDM}. 4. {Constraints} on \$f({R})\$ models from the photometric primary probes, \mn@doi{10.48550/arXiv.2409.03524}, \url {http://arxiv.org/abs/2409.03524}

\bibitem[\protect\citeauthoryear{{Euclid Collaboration: Mellier et al.}}{{Euclid Collaboration: Mellier et al.}}{2024}]{collaboration_euclid_2024}
{Euclid Collaboration: Mellier et al.} 2024, Euclid. {I}. {Overview} of the {Euclid} mission, \mn@doi{10.48550/arXiv.2405.13491}, \url {http://arxiv.org/abs/2405.13491}

\bibitem[\protect\citeauthoryear{Evrard et~al.,}{Evrard et~al.}{2002}]{evrard_galaxy_2002}
Evrard A.~E.,  et~al., 2002, \mn@doi [The Astrophysical Journal] {10.1086/340551}, 573, 7

\bibitem[\protect\citeauthoryear{Ferlito et~al.,}{Ferlito et~al.}{2023}]{ferlito_ray-tracing_2023}
Ferlito F.,  et~al., 2023, \mn@doi [Monthly Notices of the Royal Astronomical Society] {10.1093/mnras/stad2205}, 524, 5591

\bibitem[\protect\citeauthoryear{Fiorini, Koyama, Izard, Winther, Wright  \& Li}{Fiorini et~al.}{2021}]{fiorini_fast_2021}
Fiorini B.,  Koyama K.,  Izard A.,  Winther H.~A.,  Wright B.~S.,   Li B.,  2021, \mn@doi [Journal of Cosmology and Astroparticle Physics] {10.1088/1475-7516/2021/09/021}, 2021, 021

\bibitem[\protect\citeauthoryear{Fiorini, Koyama  \& Izard}{Fiorini et~al.}{2022}]{fiorini_studying_2022}
Fiorini B.,  Koyama K.,   Izard A.,  2022, \mn@doi [Journal of Cosmology and Astroparticle Physics] {10.1088/1475-7516/2022/12/028}, 2022, 028

\bibitem[\protect\citeauthoryear{Fosalba, Gaztanaga, Castander  \& Manera}{Fosalba et~al.}{2008}]{fosalba_onion_2008}
Fosalba P.,  Gaztanaga E.,  Castander F.,   Manera M.,  2008, \mn@doi [Monthly Notices of the Royal Astronomical Society] {10.1111/j.1365-2966.2008.13910.x}, 391, 435

\bibitem[\protect\citeauthoryear{Fosalba, Gaztanaga, Castander  \& Crocce}{Fosalba et~al.}{2015}]{fosalba_mice_2015}
Fosalba P.,  Gaztanaga E.,  Castander F.~J.,   Crocce M.,  2015, \mn@doi [Monthly Notices of the Royal Astronomical Society] {10.1093/mnras/stu2464}, 447, 1319

\bibitem[\protect\citeauthoryear{Frenk, White  \& Davis}{Frenk et~al.}{1983}]{frenk_nonlinear_1983}
Frenk C.~S.,  White S. D.~M.,   Davis M.,  1983, \mn@doi [The Astrophysical Journal] {10.1086/161209}, 271, 417

\bibitem[\protect\citeauthoryear{Giocoli, Baldi  \& Moscardini}{Giocoli et~al.}{2018}]{giocoli_weak_2018-1}
Giocoli C.,  Baldi M.,   Moscardini L.,  2018, \mn@doi [Monthly Notices of the Royal Astronomical Society] {10.1093/mnras/sty2465}, 481, 2813

\bibitem[\protect\citeauthoryear{Gordon, Aguiar, Rebouças, Brando, Falciano, Miranda, Koyama  \& Winther}{Gordon et~al.}{2024}]{gordon_modeling_2024}
Gordon J.,  Aguiar B. F.~d.,  Rebouças J.,  Brando G.,  Falciano F.,  Miranda V.,  Koyama K.,   Winther H.~A.,  2024, Modeling nonlinear scales with {COLA}: preparing for {LSST}-{Y1}, \mn@doi{10.48550/arXiv.2404.12344}, \url {http://arxiv.org/abs/2404.12344}

\bibitem[\protect\citeauthoryear{Gough \& Uhlemann}{Gough \& Uhlemann}{2022}]{gough_one-point_2022}
Gough A.,  Uhlemann C.,  2022, \mn@doi [Universe] {10.3390/universe8010055}, 8, 55

\bibitem[\protect\citeauthoryear{Górski, Hivon, Banday, Wandelt, Hansen, Reinecke  \& Bartelmann}{Górski et~al.}{2005}]{gorski_healpix_2005}
Górski K.~M.,  Hivon E.,  Banday A.~J.,  Wandelt B.~D.,  Hansen F.~K.,  Reinecke M.,   Bartelmann M.,  2005, \mn@doi [The Astrophysical Journal] {10.1086/427976}, 622, 759

\bibitem[\protect\citeauthoryear{Harnois-Déraps, Munshi, Valageas, Waerbeke, Brax, Coles  \& Rizzo}{Harnois-Déraps et~al.}{2015}]{harnois-deraps_testing_2015}
Harnois-Déraps J.,  Munshi D.,  Valageas P.,  Waerbeke L.~v.,  Brax P.,  Coles P.,   Rizzo L.,  2015, \mn@doi [Monthly Notices of the Royal Astronomical Society] {10.1093/mnras/stv2120}, 454, 2722

\bibitem[\protect\citeauthoryear{Harnois-Déraps, Hernandez-Aguayo, Cuesta-Lazaro, Arnold, Li, Davies  \& Cai}{Harnois-Déraps et~al.}{2022}]{harnois-deraps_mglens_2022}
Harnois-Déraps J.,  Hernandez-Aguayo C.,  Cuesta-Lazaro C.,  Arnold C.,  Li B.,  Davies C.~T.,   Cai Y.-C.,  2022, {MGLenS}: {Modified} gravity weak lensing simulations for emulation-based cosmological inference, \mn@doi{10.48550/arXiv.2211.05779}, \url {http://arxiv.org/abs/2211.05779}

\bibitem[\protect\citeauthoryear{Hartlap, Simon  \& Schneider}{Hartlap et~al.}{2007}]{hartlap_why_2007}
Hartlap J.,  Simon P.,   Schneider P.,  2007, \mn@doi [Astronomy \& Astrophysics] {10.1051/0004-6361:20066170}, 464, 399

\bibitem[\protect\citeauthoryear{Hellwing, Koyama, Bose  \& Zhao}{Hellwing et~al.}{2017}]{hellwing_revealing_2017}
Hellwing W.~A.,  Koyama K.,  Bose B.,   Zhao G.-B.,  2017, \mn@doi [Physical Review D] {10.1103/PhysRevD.96.023515}, 96, 023515

\bibitem[\protect\citeauthoryear{Higuchi \& Shirasaki}{Higuchi \& Shirasaki}{2016}]{higuchi_imprint_2016}
Higuchi Y.,  Shirasaki M.,  2016, The imprint of \$f({R})\$ gravity on weak gravitational lensing {I}: {Connection} between observables and large scale structure, \url {http://arxiv.org/abs/1603.01325}

\bibitem[\protect\citeauthoryear{Hilbert, Hartlap, White  \& Schneider}{Hilbert et~al.}{2009}]{hilbert_ray-tracing_2009}
Hilbert S.,  Hartlap J.,  White S. D.~M.,   Schneider P.,  2009, \mn@doi [Astronomy \& Astrophysics] {10.1051/0004-6361/200811054}, 499, 31

\bibitem[\protect\citeauthoryear{Hinshaw et~al.,}{Hinshaw et~al.}{2013}]{hinshaw_nine-year_2013}
Hinshaw G.,  et~al., 2013, Nine-{Year} {Wilkinson} {Microwave} {Anisotropy} {Probe} ({WMAP}) {Observations}: {Cosmological} {Parameter} {Results}, \url {http://arxiv.org/abs/1212.5226}

\bibitem[\protect\citeauthoryear{Hu \& Sawicki}{Hu \& Sawicki}{2007}]{hu_models_2007}
Hu W.,  Sawicki I.,  2007, \mn@doi [Physical Review D] {10.1103/PhysRevD.76.064004}, 76, 064004

\bibitem[\protect\citeauthoryear{Ivezi{\'c} et~al.,}{Ivezi{\'c} et~al.}{2019}]{ivezic_lsst_2019}
Ivezi{\'c} {\v{Z}}.,  et~al., 2019, \mn@doi [The Astrophysical Journal] {10.3847/1538-4357/ab042c}, 873, 111

\bibitem[\protect\citeauthoryear{Izard, Crocce  \& Fosalba}{Izard et~al.}{2016}]{izard_ice-cola_2016}
Izard A.,  Crocce M.,   Fosalba P.,  2016, \mn@doi [Monthly Notices of the Royal Astronomical Society] {10.1093/mnras/stw797}, 459, 2327

\bibitem[\protect\citeauthoryear{Izard, Fosalba  \& Crocce}{Izard et~al.}{2018}]{izard_ice-cola_2018}
Izard A.,  Fosalba P.,   Crocce M.,  2018, \mn@doi [Monthly Notices of the Royal Astronomical Society] {10.1093/mnras/stx2544}, 473, 3051

\bibitem[\protect\citeauthoryear{Jiang, Liu, Li, Barrera-Hinojosa, Zhang  \& Fang}{Jiang et~al.}{2024}]{jiang_minkowski_2024}
Jiang A.,  Liu W.,  Li B.,  Barrera-Hinojosa C.,  Zhang Y.,   Fang W.,  2024, Minkowski {Functionals} of {Large}-{Scale} {Structure} as a {Probe} of {Modified} {Gravity}, \url {http://arxiv.org/abs/2305.04520}

\bibitem[\protect\citeauthoryear{Kacprzak et~al.,}{Kacprzak et~al.}{2016}]{kacprzak_cosmology_2016}
Kacprzak T.,  et~al., 2016, \mn@doi [Monthly Notices of the Royal Astronomical Society] {10.1093/mnras/stw2070}, 463, 3653

\bibitem[\protect\citeauthoryear{Khoury \& Weltman}{Khoury \& Weltman}{2004}]{khoury_chameleon_2004}
Khoury J.,  Weltman A.,  2004, \mn@doi [Physical Review Letters] {10.1103/PhysRevLett.93.171104}, 93, 171104

\bibitem[\protect\citeauthoryear{Koyama}{Koyama}{2016}]{koyama_cosmological_2016}
Koyama K.,  2016, \mn@doi [Reports on Progress in Physics] {10.1088/0034-4885/79/4/046902}, 79, 046902

\bibitem[\protect\citeauthoryear{Kratochvil, Haiman  \& May}{Kratochvil et~al.}{2009}]{kratochvil_probing_2009}
Kratochvil J.~M.,  Haiman Z.,   May M.,  2009, Probing {Cosmology} with {Weak} {Lensing} {Peak} {Counts}, \url {http://arxiv.org/abs/0907.0486}

\bibitem[\protect\citeauthoryear{{LIGO Scientific Collaboration}, {Virgo Collaboration}, Monitor  \& INTEGRAL}{{LIGO Scientific Collaboration} et~al.}{2017}]{collaboration_gravitational_2017}
{LIGO Scientific Collaboration} {Virgo Collaboration} Monitor F. G.-R.~B.,   INTEGRAL 2017, \mn@doi [The Astrophysical Journal Letters] {10.3847/2041-8213/aa920c}, 848, L13

\bibitem[\protect\citeauthoryear{Leclercq \& Heavens}{Leclercq \& Heavens}{2021}]{leclercq_accuracy_2021}
Leclercq F.,  Heavens A.,  2021, \mn@doi [Monthly Notices of the Royal Astronomical Society: Letters] {10.1093/mnrasl/slab081}, 506, L85

\bibitem[\protect\citeauthoryear{Lewis, Challinor  \& Lasenby}{Lewis et~al.}{2000}]{lewis_efficient_2000}
Lewis A.,  Challinor A.,   Lasenby A.,  2000, \mn@doi [The Astrophysical Journal] {10.1086/309179}, 538, 473

\bibitem[\protect\citeauthoryear{Lewis et~al.,}{Lewis et~al.}{2019}]{antony_lewis_2019_3452064}
Lewis A.,  et~al., 2019, cmbant/CAMB: 1.0.8, \mn@doi{10.5281/zenodo.3452064}, \url {https://doi.org/10.5281/zenodo.3452064}

\bibitem[\protect\citeauthoryear{Ling, Wang, Li, Li, Wang  \& Gao}{Ling et~al.}{2015}]{ling_distinguishing_2015}
Ling C.,  Wang Q.,  Li R.,  Li B.,  Wang J.,   Gao L.,  2015, \mn@doi [Physical Review D] {10.1103/PhysRevD.92.064024}, 92, 064024

\bibitem[\protect\citeauthoryear{List \& Hahn}{List \& Hahn}{2024}]{list_perturbation-theory_2024}
List F.,  Hahn O.,  2024, Perturbation-theory informed integrators for cosmological simulations, \mn@doi{10.48550/arXiv.2301.09655}, \url {http://arxiv.org/abs/2301.09655}

\bibitem[\protect\citeauthoryear{Liu et~al.,}{Liu et~al.}{2016}]{liu_constraining_2016}
Liu X.,  et~al., 2016, \mn@doi [Physical Review Letters] {10.1103/PhysRevLett.117.051101}, 117, 051101

\bibitem[\protect\citeauthoryear{Mancini \& Bose}{Mancini \& Bose}{2023}]{mancini_degeneracies_2023}
Mancini A.~S.,  Bose B.,  2023, \mn@doi [The Open Journal of Astrophysics] {10.21105/astro.2305.06350}, 6, 10.21105/astro.2305.06350

\bibitem[\protect\citeauthoryear{Mancini, Lin  \& McEwen}{Mancini et~al.}{2024}]{mancini_field-level_2024}
Mancini A.~S.,  Lin K.,   McEwen J.~D.,  2024, Field-level cosmological model selection: field-level simulation-based inference for {Stage} {IV} cosmic shear can distinguish dynamical dark energy, \mn@doi{10.48550/arXiv.2410.10616}, \url {http://arxiv.org/abs/2410.10616}

\bibitem[\protect\citeauthoryear{Marian, Smith  \& Bernstein}{Marian et~al.}{2009}]{marian_cosmology_2009}
Marian L.,  Smith R.~E.,   Bernstein G.~M.,  2009, \mn@doi [The Astrophysical Journal] {10.1088/0004-637X/698/1/L33}, 698, L33

\bibitem[\protect\citeauthoryear{Martinet, Bartlett, Kiessling  \& Sartoris}{Martinet et~al.}{2015}]{martinet_constraining_2015}
Martinet N.,  Bartlett J.~G.,  Kiessling A.,   Sartoris B.,  2015, \mn@doi [Astronomy \& Astrophysics] {10.1051/0004-6361/201425164}, 581, A101

\bibitem[\protect\citeauthoryear{Martinet et~al.,}{Martinet et~al.}{2018}]{martinet_kids-450_2018}
Martinet N.,  et~al., 2018, \mn@doi [Monthly Notices of the Royal Astronomical Society] {10.1093/mnras/stx2793}, 474, 712

\bibitem[\protect\citeauthoryear{McEwen, Fang, Hirata  \& Blazek}{McEwen et~al.}{2016}]{mcewen_fast-pt_2016}
McEwen J.~E.,  Fang X.,  Hirata C.~M.,   Blazek J.~A.,  2016, \mn@doi [Journal of Cosmology and Astroparticle Physics] {10.1088/1475-7516/2016/09/015}, 2016, 015

\bibitem[\protect\citeauthoryear{Mead, Peacock, Heymans, Joudaki  \& Heavens}{Mead et~al.}{2015}]{mead_accurate_2015}
Mead A.,  Peacock J.,  Heymans C.,  Joudaki S.,   Heavens A.,  2015, \mn@doi [Monthly Notices of the Royal Astronomical Society] {10.1093/mnras/stv2036}, 454, 1958

\bibitem[\protect\citeauthoryear{Mecke, Buchert  \& Wagner}{Mecke et~al.}{1993}]{mecke_robust_1993}
Mecke K.~R.,  Buchert T.,   Wagner H.,  1993, Robust {Morphological} {Measures} for {Large}-{Scale} {Structure} in the {Universe}, \mn@doi{10.48550/arXiv.astro-ph/9312028}, \url {http://arxiv.org/abs/astro-ph/9312028}

\bibitem[\protect\citeauthoryear{Munshi, Pratten, Valageas, Coles  \& Brax}{Munshi et~al.}{2016}]{munshi_galaxy_2016}
Munshi D.,  Pratten G.,  Valageas P.,  Coles P.,   Brax P.,  2016, \mn@doi [Monthly Notices of the Royal Astronomical Society] {10.1093/mnras/stv2724}, 456, 1627

\bibitem[\protect\citeauthoryear{Munshi, Namikawa, Kitching, McEwen, Takahashi, Bouchet, Taruya  \& Bose}{Munshi et~al.}{2020}]{munshi_weak_2020}
Munshi D.,  Namikawa T.,  Kitching T.~D.,  McEwen J.~D.,  Takahashi R.,  Bouchet F.~R.,  Taruya A.,   Bose B.,  2020, \mn@doi [Monthly Notices of the Royal Astronomical Society] {10.1093/mnras/staa296}, 493, 3985

\bibitem[\protect\citeauthoryear{Parroni, Cardone, Maoli  \& Scaramella}{Parroni et~al.}{2020}]{parroni_going_2020}
Parroni C.,  Cardone V.~F.,  Maoli R.,   Scaramella R.,  2020, \mn@doi [Astronomy \& Astrophysics] {10.1051/0004-6361/201935988}, 633, A71

\bibitem[\protect\citeauthoryear{Petri}{Petri}{2016}]{petri_mocking_2016}
Petri A.,  2016, Mocking the {Weak} {Lensing} universe: the {LensTools} python computing package, \url {http://arxiv.org/abs/1606.01903}

\bibitem[\protect\citeauthoryear{Petri, Liu, Haiman, May, Hui  \& Kratochvil}{Petri et~al.}{2015}]{petri_emulating_2015}
Petri A.,  Liu J.,  Haiman Z.,  May M.,  Hui L.,   Kratochvil J.~M.,  2015, \mn@doi [Physical Review D] {10.1103/PhysRevD.91.103511}, 91, 103511

\bibitem[\protect\citeauthoryear{Philcox}{Philcox}{2023}]{philcox_optimal_2023}
Philcox O. H.~E.,  2023, Optimal {Estimation} of the {Binned} {Mask}-{Free} {Power} {Spectrum}, {Bispectrum}, and {Trispectrum} on the {Full} {Sky}: {Scalar} {Edition}, \url {http://arxiv.org/abs/2303.08828}

\bibitem[\protect\citeauthoryear{Pontzen, Slosar, Roth  \& Peiris}{Pontzen et~al.}{2016}]{pontzen_inverted_2016}
Pontzen A.,  Slosar A.,  Roth N.,   Peiris H.~V.,  2016, \mn@doi [Physical Review D] {10.1103/PhysRevD.93.103519}, 93, 103519

\bibitem[\protect\citeauthoryear{Porqueres, Heavens, Mortlock, Lavaux  \& Makinen}{Porqueres et~al.}{2023}]{porqueres_field-level_2023}
Porqueres N.,  Heavens A.,  Mortlock D.,  Lavaux G.,   Makinen T.~L.,  2023, Field-level inference of cosmic shear with intrinsic alignments and baryons, \mn@doi{10.48550/arXiv.2304.04785}, \url {http://arxiv.org/abs/2304.04785}

\bibitem[\protect\citeauthoryear{Schmalzing, Kerscher  \& Buchert}{Schmalzing et~al.}{1995}]{schmalzing_minkowski_1995}
Schmalzing J.,  Kerscher M.,   Buchert T.,  1995, Minkowski {Functionals} in {Cosmology}, \mn@doi{10.48550/arXiv.astro-ph/9508154}, \url {http://arxiv.org/abs/astro-ph/9508154}

\bibitem[\protect\citeauthoryear{Schneider et~al.,}{Schneider et~al.}{2020}]{schneider_baryonic_2020}
Schneider A.,  et~al., 2020, \mn@doi [Journal of Cosmology and Astroparticle Physics] {10.1088/1475-7516/2020/04/020}, 2020, 020

\bibitem[\protect\citeauthoryear{Scoccimarro, Couchman  \& Frieman}{Scoccimarro et~al.}{1999}]{scoccimarro_bispectrum_1999}
Scoccimarro R.,  Couchman H. M.~P.,   Frieman J.~A.,  1999, \mn@doi [The Astrophysical Journal] {10.1086/307220}, 517, 531

\bibitem[\protect\citeauthoryear{Sefusatti, Crocce, Scoccimarro  \& Couchman}{Sefusatti et~al.}{2016}]{sefusatti_accurate_2016}
Sefusatti E.,  Crocce M.,  Scoccimarro R.,   Couchman H.,  2016, \mn@doi [Monthly Notices of the Royal Astronomical Society] {10.1093/mnras/stw1229}, 460, 3624

\bibitem[\protect\citeauthoryear{Shandarin \& Zeldovich}{Shandarin \& Zeldovich}{1989}]{shandarin_large-scale_1989}
Shandarin S.~F.,  Zeldovich Y.~B.,  1989, \mn@doi [Reviews of Modern Physics] {10.1103/RevModPhys.61.185}, 61, 185

\bibitem[\protect\citeauthoryear{Shirasaki, Hamana  \& Yoshida}{Shirasaki et~al.}{2015}]{shirasaki_probing_2015}
Shirasaki M.,  Hamana T.,   Yoshida N.,  2015, \mn@doi [Monthly Notices of the Royal Astronomical Society] {10.1093/mnras/stv1854}, 453, 3044

\bibitem[\protect\citeauthoryear{Shirasaki, Nishimichi, Li  \& Higuchi}{Shirasaki et~al.}{2017}]{shirasaki_imprint_2017}
Shirasaki M.,  Nishimichi T.,  Li B.,   Higuchi Y.,  2017, \mn@doi [Monthly Notices of the Royal Astronomical Society] {10.1093/mnras/stw3254}, 466, 2402

\bibitem[\protect\citeauthoryear{Springel}{Springel}{2005}]{springel_cosmological_2005}
Springel V.,  2005, \mn@doi [Monthly Notices of the Royal Astronomical Society] {10.1111/j.1365-2966.2005.09655.x}, 364, 1105

\bibitem[\protect\citeauthoryear{Sáez-Casares, Rasera  \& Li}{Sáez-Casares et~al.}{2023}]{saez-casares_e-mantis_2023}
Sáez-Casares I.,  Rasera Y.,   Li B.,  2023, The e-{MANTIS} emulator: fast predictions of the non-linear matter power spectrum in \$f({R})\${CDM} cosmology, \url {http://arxiv.org/abs/2303.08899}

\bibitem[\protect\citeauthoryear{Takahashi, Hamana, Shirasaki, Namikawa, Nishimichi, Osato  \& Shiroyama}{Takahashi et~al.}{2017}]{takahashi_full-sky_2017}
Takahashi R.,  Hamana T.,  Shirasaki M.,  Namikawa T.,  Nishimichi T.,  Osato K.,   Shiroyama K.,  2017, Full-sky {Gravitational} {Lensing} {Simulation} for {Large}-area {Galaxy} {Surveys} and {Cosmic} {Microwave} {Background} {Experiments}, \url {http://arxiv.org/abs/1706.01472}

\bibitem[\protect\citeauthoryear{Tassev, Zaldarriaga  \& Eisenstein}{Tassev et~al.}{2013}]{tassev_solving_2013}
Tassev S.,  Zaldarriaga M.,   Eisenstein D.,  2013, \mn@doi [Journal of Cosmology and Astroparticle Physics] {10.1088/1475-7516/2013/06/036}, 2013, 036

\bibitem[\protect\citeauthoryear{Tassev, Eisenstein, Wandelt  \& Zaldarriaga}{Tassev et~al.}{2015}]{tassev_scola_2015}
Tassev S.,  Eisenstein D.~J.,  Wandelt B.~D.,   Zaldarriaga M.,  2015, {sCOLA}: {The} {N}-body {COLA} {Method} {Extended} to the {Spatial} {Domain}, \mn@doi{10.48550/arXiv.1502.07751}, \url {http://arxiv.org/abs/1502.07751}

\bibitem[\protect\citeauthoryear{Thiele, Hill  \& Smith}{Thiele et~al.}{2020}]{thiele_accurate_2020}
Thiele L.,  Hill J.~C.,   Smith K.~M.,  2020, \mn@doi [Physical Review D] {10.1103/PhysRevD.102.123545}, 102, 123545

\bibitem[\protect\citeauthoryear{Tsedrik, Bose, Carrilho, Pourtsidou, Pamuk, Casas  \& Lesgourgues}{Tsedrik et~al.}{2024}]{tsedrik_stage-iv_2024}
Tsedrik M.,  Bose B.,  Carrilho P.,  Pourtsidou A.,  Pamuk S.,  Casas S.,   Lesgourgues J.,  2024, Stage-{IV} {Cosmic} {Shear} with {Modified} {Gravity} and {Model}-independent {Screening}, \mn@doi{10.48550/arXiv.2404.11508}, \url {http://arxiv.org/abs/2404.11508}

\bibitem[\protect\citeauthoryear{Valentino et~al.,}{Valentino et~al.}{2021}]{valentino_realm_2021}
Valentino E.~D.,  et~al., 2021, \mn@doi [Classical and Quantum Gravity] {10.1088/1361-6382/ac086d}, 38, 153001

\bibitem[\protect\citeauthoryear{Weinberg, Mortonson, Eisenstein, Hirata, Riess  \& Rozo}{Weinberg et~al.}{2013}]{weinberg_observational_2013}
Weinberg D.~H.,  Mortonson M.~J.,  Eisenstein D.~J.,  Hirata C.,  Riess A.~G.,   Rozo E.,  2013, \mn@doi [Physics Reports] {10.1016/j.physrep.2013.05.001}, 530, 87

\bibitem[\protect\citeauthoryear{Will}{Will}{2014}]{will_confrontation_2014}
Will C.~M.,  2014, \mn@doi [Living Reviews in Relativity] {10.12942/lrr-2014-4}, 17, 4

\bibitem[\protect\citeauthoryear{Winther, Koyama, Manera, Wright  \& Zhao}{Winther et~al.}{2017}]{winther_cola_2017}
Winther H.~A.,  Koyama K.,  Manera M.,  Wright B.~S.,   Zhao G.-B.,  2017, \mn@doi [Journal of Cosmology and Astroparticle Physics] {10.1088/1475-7516/2017/08/006}, 2017, 006

\bibitem[\protect\citeauthoryear{Wright, Winther  \& Koyama}{Wright et~al.}{2017}]{wright_cola_2017}
Wright B.~S.,  Winther H.~A.,   Koyama K.,  2017, \mn@doi [Journal of Cosmology and Astroparticle Physics] {10.1088/1475-7516/2017/10/054}, 2017, 054

\bibitem[\protect\citeauthoryear{Wright, Gupta, Baker, Valogiannis  \& Fiorini}{Wright et~al.}{2023}]{wright_texttthi-cola_2023}
Wright B.~S.,  Gupta A.~S.,  Baker T.,  Valogiannis G.,   Fiorini B.,  2023, \mn@doi [Journal of Cosmology and Astroparticle Physics] {10.1088/1475-7516/2023/03/040}, 2023, 040

\bibitem[\protect\citeauthoryear{Zhou, Li, Dodelson  \& Mandelbaum}{Zhou et~al.}{2023}]{zhou_accurate_2023}
Zhou A.~J.,  Li X.,  Dodelson S.,   Mandelbaum R.,  2023, Accurate field-level weak lensing inference for precision cosmology, \url {http://arxiv.org/abs/2312.08934}

\makeatother
\end{thebibliography}

\appendix

\bsp	
\label{lastpage}
\end{document}